\documentclass[journal,comsoc]{IEEEtran}

\usepackage{amsmath,amssymb,amsfonts}
\usepackage{color, soul}
\usepackage{float}
\usepackage{caption} 
\usepackage[utf8]{inputenc}
\usepackage[english]{babel}
\usepackage[acronym]{glossaries}
\makeglossaries

\newacronym{ai}{AI}{Artificial Intelligence}
\newacronym{ml}{ML}{Machine Learning}
\newacronym{edc}{EDC}{Eclipse Dataspace Components}
\newacronym{llm}{LLM}{Large Language Model}
\newacronym{rag}{RAG}{Retrieval Augmented Generation}
\newacronym{mvds}{MVDS}{Minimal Viable Data Space}
\newacronym{mcp}{MCP}{Model Context Protocol}
\newacronym{tui}{TUI}{Terminal User Interface}
\newacronym{api}{API}{Application Programming Interface}
\newacronym{sla}{SLA}{Service Level Agreement}
\newacronym{sml}{SML}{Small Language Model}
\newacronym{marl}{MARL}{multi-agent reinforcement learning}
\newacronym{vm}{VM}{Virtual Machine}
\newacronym{iaas}{IaaS}{Infrastructure as a Service}
\newacronym{cpu}{CPU}{Central Processing Unit}
\newacronym{iot}{IoT}{Internet of Things}
\newacronym{cnn}{CNN}{Convolutional Neural Network}
\newacronym{mlp}{MLP}{Multi-Layer Perceptron}
\newacronym{rpi}{RPI}{Raspberry Pi}
\newacronym{genai}{GenAI}{Generative artificial intelligence}
\newacronym{mape}{MAPE}{Mean Absolute Percentage Error}
\usepackage{paralist}
\usepackage{algorithmic}
\usepackage{graphicx}
\usepackage{textcomp}
\usepackage{xcolor}
\usepackage{comment}
\usepackage{graphicx}
\usepackage{subcaption}
\usepackage[ruled,vlined,linesnumbered]{algorithm2e}
\usepackage{booktabs}
\usepackage{multirow}
\usepackage{tabularx}
\def\BibTeX{{\rm B\kern-.05em{\sc i\kern-.025em b}\kern-.08em
    T\kern-.1667em\lower.7ex\hbox{E}\kern-.125emX}}

\newcommand{\eg}{e.\,g.\ }
\usepackage{pdfcomment}

\makeatletter
\@ifundefined{missingfigure}{}{}
\@ifundefined{textcomment}{}{}
\@ifundefined{sidecomment}{}{}
\@ifundefined{todo}{}{}
\@ifundefined{TODO}{}{}
\@ifundefined{todofix}{}{}
\@ifundefined{change}{}{}
\makeatother

\begin{document}

\title{A Self-Calibrating Agentic AI Framework for Autonomous Edge Resource Allocation
}

\author{
  \IEEEauthorblockN{Fin Gentzen, Marla Grunewald, Iulisloi Zacarias, Mounir Bensalem and Admela Jukan} \\
  \IEEEauthorblockA{Institut f\"ur Datentechnik und Kommunikationsnetze\\
  Technische Universit\"at Braunschweig, Germany\\
  Email: \{f.gentzen, marla.grunewald, i.zacarias, mounir.bensalem,  a.jukan\}@tu-bs.de}
}

\maketitle

\begin{abstract}
Large Language Models (LLMs) are increasingly deployed as autonomous agents, transitioning from static conversational interfaces to dynamic systems capable of complex reasoning, tool execution, and decision-making. However, the operational reliability of these agentic AI systems is fundamentally challenged by the absence of reliable ground truth in open-ended environments and the risk of increasing operational drift over time. To address this challenge, we propose and experimentally evaluate an agentic AI framework, designed to enforce autonomous integrity within LLM-driven systems. We design a self-calibration mechanism that mitigates drift and dynamically approximates ground truth by incorporating an ARIMA forecaster, without requiring continuous human oversight. To demonstrate the effectiveness and reliability of our methodology, we apply it to the complex domain of profiling the resource usage of zero-knowledge workloads in edge computing networks. Experimental results show that the proposed self-calibrating agentic framework successfully profiles the zero-knowledge workloads, achieving a higher accuracy than baseline LLM agents by 91.7\% for resource usage prediction and improving the prediction speed by 71.7\% compared to pure profiling, establishing a robust foundation for deploying autonomous AI in  decentralized infrastructures. Furthermore, the ground truth generation using the proposed ARIMA leaping algorithm is 52\% faster than a standard ARIMA forecasting algorithm, while achieving the same accuracy.

\end{abstract}

\begin{IEEEkeywords}
Agentic AI, Self-Calibration, Edge Network, AI Workload
\end{IEEEkeywords}

\section{Introduction}

\IEEEPARstart{T}{he} exponential growth of distributed \gls{ai} applications is forcing network service providers to rethink their network architectures and deployment plans. \Gls{ai} training and inference processes, once centralized in large data centers, are increasingly migrating to the network edge. Placing \Gls{ai} workloads at the network's edge offers compelling advantages for time-sensitive and privacy-concerned applications.  Especially lightweight \gls{ai} workloads are expected to be placed closer to the end-user, thereby distributing the computing tasks along the computing continuum, whereas resource-hungry workloads are still likely to be placed in data centers. For an efficient implementation of the compute continuum, edge devices are inherently resource-constrained devices in a highly heterogeneous landscape~\cite{zhou2019edge} and, as such, present a challenge for efficient workload placement. To address this challenge, the rapid evolution of autonomous agentic \gls{ai} is giving rise to agentic edge intelligence~\cite{10.1145/3773274.3777421}.

Despite significant benefits, agentic edge intelligence exposes a fundamental operational problem. In contrast to homogeneous cloud data centers, where elastic resource pools (e.g., CPU, GPU, memory, and storage) are available, (far-) edge environments are characterized by hardware heterogeneity, limited computational capacity, and constrained memory and storage. Correctly allocating resources for \gls{ai} workloads across the computing continuum requires precise knowledge of resource consumption in advance, which is typically unavailable. In edge environments, resource consumption can drift during execution, for example, by requiring fewer or more computing resources compared to the initial estimation; in one case it can lead to resource underutilization, while in the other, the \gls{ai} workload might fail, by abruptly finishing its execution without producing any results. The workload would then likely need to be reallocated to computing nodes capable of handling it, and processing would restart. This situation may lead to \glspl{sla} breaches due to increased processing latency and low user satisfaction, and potentially disastrous outcomes in time-sensitive systems. What is needed is a precise estimation of the resource consumption of \gls{ai} workloads, backed by the reliable ground-truth data to validate estimation algorithms and as well as self-calibration mechanisms to counteract runtime drift.

This paper proposes a novel self-calibration agentic AI framework that can efficiently address the problem of edge resource allocation for \gls{ai} workloads with zero-knowledge \emph{a priori}. It assumes that \gls{ai} workloads arrive at edge devices as black-box executables paired with a dataset which may vary in size and format. Upon arrival, no accompanying metadata describing their memory footprint, CPU and GPU requirements, or execution time is available. Our solution is autonomous, as it does not require human oversight while it combines \gls{llm}-based prediction with empirical telemetry observation, dynamically approximating ground truth. An AI Workload Agent orchestrates four modules: i) LLM Zero-Shot Estimation that evaluates and predicts the workload resource usage by performing static code analysis; ii) Active Profiling, that combines partial empirical evaluation of the workloads with an ARIMA-based forecasting engine to extrapolate the collected measurements to predict full-scale resource consumption; iii) \gls{llm}~+~\gls{rag} Estimation, that uses previous knowledge generated by the system to predict system load with higher accuracy without empirical evaluation of the workload; and iv) Re-Profile and Calibrate module that is triggered when the system fails to fulfill any of the environment constraints. 
\par The novel contributions of this study can be summarized as follows:

\begin{itemize}
    \item A self-calibrating agentic AI architecture for zero-knowledge workload profiling, integrating a reasoning controller, a policy and constraints manager, and four complementary profiling modules that can be invoked independently or concurrently. To the best of our knowledge, this is the first framework that autonomously generates and refines its own ground truth for edge resource allocation.

    \item A formal system model for bootstrap profiling, covering both discrete-memory and unified-memory architectures, which employs $M/G/\infty$ queuing for CPU modeling, occupancy-based GPU modeling, and ARIMA-based forecasting to extrapolate sandbox executions to full-scale resource footprints.

    \item An adaptive parameter search algorithm with three leaping strategies that navigates the sample/epoch configuration space under strict time budgets, terminating early once forecast confidence exceeds validated thresholds.

    \item An experimental framework with open benchmark dataset of 53 profiled AI workloads spanning eight model architectures, six datasets, and three heterogeneous hardware platforms (Raspberry Pi 5, NVIDIA Jetson Thor, GPU workstation), released to support full reproducibility of research and for usage of data in the research community. 

  \item A comprehensive experimental evaluation shows how our proposed agentic framework reduces prediction error from over 200\% MAPE to single-digit MAPE for well-covered workload classes, while it is still faster than classical ARIMA-only workload estimators. 
       
\end{itemize}

The rest of the paper is organized as follows. Section II presents related work. Section~\ref{sec:arch} describes the proposed architecture, illustrating the hierarchical agentic \gls{ai} workflow, the developed resource prediction workflows, and the tools employed. Section~\ref{sec:system-model} presents a theoretical model of the \textit{Bootstrap Profiling} method together with the problem formulation. In Section ~\ref{sec:data}, we present the dataset that was created in this study. Section~\ref{sec:eval} presents and discusses the analytical and experimental results. Section~\ref{sec:concl} concludes the paper and provides directions for further research. 
\section{Related Work}\label{sec:relwork}

\subsection{LLMs and Agentic AI in Network Management}

Industry forums have extensively demonstrated how \gls{genai} reshaped network monitoring and orchestration \cite{francois2025-nmrg, yudong2025, 10891637}. Moving beyond basic conversational interfaces, these models translate high-level intents into dynamic network configurations \cite{asif2025, provvedi2026, dzeparoska2025}. Furthermore, autonomous entities have been embedded to mitigate network outages \cite{10891042}, and modern frameworks employ \gls{rag} to bind generated parameters to external verifiable knowledge bases, significantly reducing model hallucinations \cite{dzeparoska2024, 11303308}.

In line with these trends, our framework relies on the complex reasoning capabilities of \glspl{llm} and \gls{rag} to orchestrate tasks and translate high-level constraints autonomously. However, our operational assumptions differ significantly. Most existing applications assume centralized environments with virtually unbounded computing power or rely on infrastructure knowledge that is confined to explicitly declared models, such as YANG or TOSCA. What these papers are not addressing is the challenge of orchestrating zero-knowledge \gls{ai} executables arriving at highly constrained and heterogeneous edge architectures without any accompanying metadata. We directly address this gap by proposing a self-calibrating agentic \gls{ai} architecture that integrates a reasoning controller to autonomously generate and refine its own ground truth without human oversight.

\subsection{Theoretical Modeling of AI Telemetry Data}

To correctly allocate resources across the computing continuum, establishing an analytical foundation for resource consumption is critical. Earlier works have proposed theoretical abstractions to predict hardware loads. For instance, statistical queue models have been successfully utilized for reliable forecasting of future CPU consumption based on request arrival patterns \cite{hammer2018queue}. Similarly, in the domain of hardware acceleration, studies have formalized the analysis of resource utilization and occupancy parameters on GPUs \cite{pimple2019}. 

Based on these studies, however applied in different contexts, we apply mathematical queueing and occupancy theories to abstract and estimate physical hardware utilization during task execution. Our methodology extends the existing literature majorly, by integrating both assumptions together with constraints for memory, and using this combination to describe a mathematical telemetry benchmark system for AI workloads. 

\subsection{Telemetry-Driven Workload Forecasting and Profiling Models, and Reproducibility}

Deploying workloads accurately across the edge continuum requires anticipating resource consumption, a challenge often tackled through statistical forecasting and deep learning. Traditional machine learning models have been successfully used to predict virtual machine execution times and to model network interference impacts \cite{10.1145/3772052.3772251, 11150566}. For extrapolating short-term telemetry observations, time series forecasting methods such as ARIMA are commonly deployed \cite{10466718, 6881647}. More recently, deep learning ensembles and lightweight feedforward neural networks have been developed to capture temporal and spatial patterns for predicting execution times across edge-fog-cloud architectures \cite{10837478, 11416223}. Active telemetry profiling is also widely utilized to evaluate multi-criteria cost functions dynamically and swap between complex and lightweight models to prevent hardware overload \cite{10948450}.

We operate similarly by relying on active profiling and ARIMA-based forecasting engines to extrapolate partial empirical observations into full-scale predictions. The key difference is that existing methods are relatively slow and need to be fed a lot of data to make sophisticated forecasts. They cannot speed up the forecasting process based on dynamic accuracy changes by filtering the forecasting data.
We overcome this limitation by developing a leaping strategy that significantly speeds up the ARIMA forecasting time, while achieving a very comparable accuracy.

A review of the literature reveals a distinct lack of comprehensive open datasets for evaluating autonomous edge resource allocation. Most related frameworks validate their estimation algorithms using proprietary network traces, constrained simulations, or legacy datasets that do not reflect the diverse operational drift of modern containerized AI applications on physical edge hardware.

While we perform rigorous experimental evaluations similarly to the existing literature, our approach differs by prioritizing full transparency and physical hardware heterogeneity across the compute continuum. The existing literature does not address the community need for reliable ground-truth telemetry spanning multiple modern model architectures and diverse physical edge devices. We explicitly address this missing literature in our fourth contribution by releasing an experimental framework and an open benchmark dataset of 53 profiled \gls{ai} workloads spanning eight model architectures, six datasets, and three heterogeneous hardware platforms, thereby supporting the full reproducibility of research in agentic edge intelligence.

\subsection{Edge Resource Orchestration and Multi-Agent Systems}

To manage the limited capacity and significant hardware heterogeneity of the edge continuum, \gls{marl} is frequently used for task offloading and resource scheduling \cite{10465255, hady2025multi, yao2025multi}. In these systems, distributed agents optimize policies for dynamic load balancing \cite{liu2025dynamic}, and digital twins are increasingly used to assist in decentralized environments \cite{11224639}. To overcome the inflexibility of pure reinforcement learning, newer hybrid architectures combine classical optimization techniques with \gls{llm}-based reasoning for adaptive agent placement \cite{11373977} and serverless distributed inference scaling \cite{11479931}.

While our framework shares the ultimate goal of minimizing \glspl{sla} breaches in dynamic edge environments, we differ fundamentally in our problem formulation and operational constraints. Existing \gls{marl} approaches function primarily as decision-making policy engines for the resource allocation process. These systems assume that the environmental state, such as the resource footprint of a workload, can be observed or learned over many episodes. Consequently, they suffer from a severe cold-start problem and require extensive offline pre-training, rendering them ineffective when confronted with entirely unseen, black-box executables. Furthermore, when workloads experience runtime drift, \gls{marl} agents treat this as environmental non-stationarity, necessitating slow and computationally expensive online re-convergence that violates strict edge time budgets. Because these works focus on the placement policy rather than the precursor problem of zero-knowledge resource profiling, they serve as a different class of system. We directly address this estimation bottleneck by proving that our agentic architecture shortening the cold-start phase severely, generating highly accurate resource forecasts under strict time budgets while having no initial information about the workloads at all.
Our solution can be used by other \gls{marl} approaches to enrich the state information and make a sophisticated resource allocation decision. 
\section{Architecture}
\label{sec:arch}
\begin{figure*}[tbp]
    \centering
    \includegraphics[width=1\linewidth]{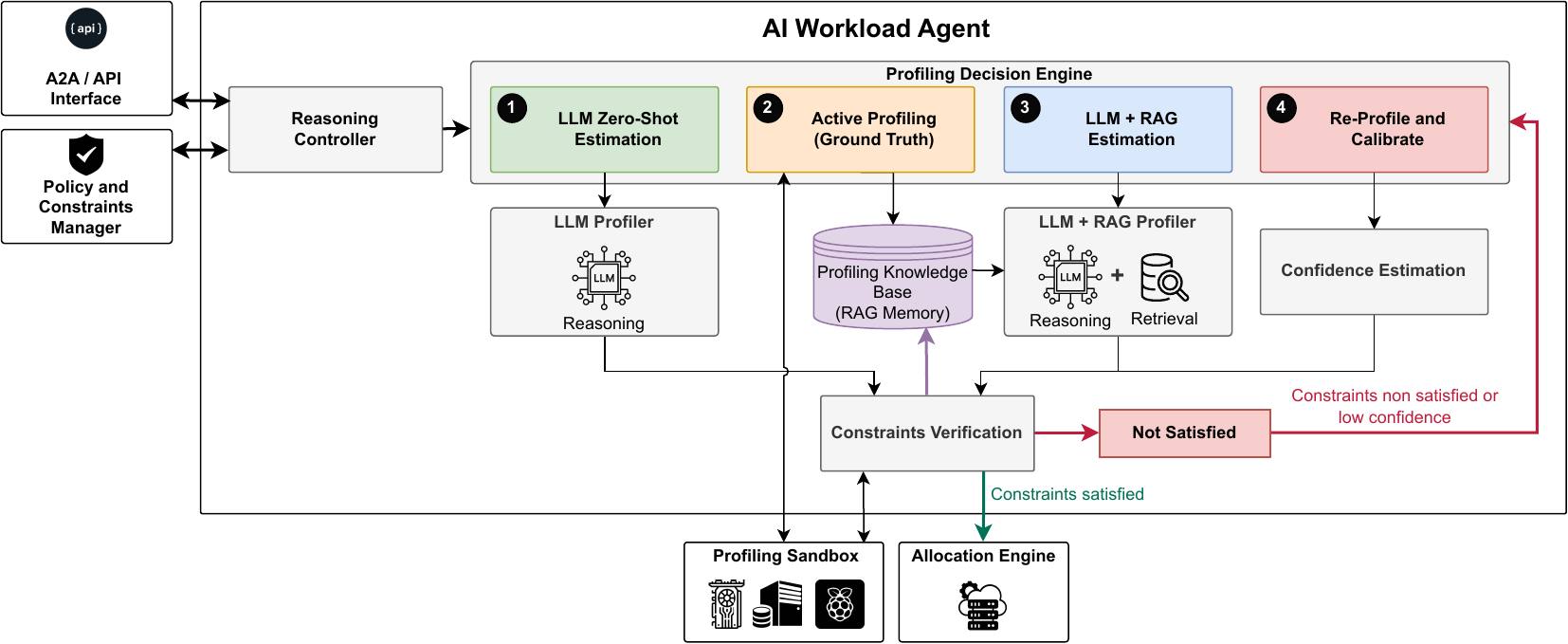}
    \caption{System workflow architecture}
    \label{fig:Arch}
\end{figure*}

\figurename~\ref{fig:Arch} shows the proposed architecture. It includes an agent, the AI Workload Agent, consisting of an A2A Interface, a Reasoning \gls{llm} and four different main modules, that can be called independently or concurrently by the \gls{llm}. The final output of this AI Workload Agent is the predicted resource usage of an (AI) workload in a distributed edge cluster. In our implementation, an (AI) workload is provided to the system in JSON format, pointing to the Python file in which the AI task is described and a pointer to the dataset that is used for the training process. Together with the JSON file, the reasoning controller is connected to the Policy and Constrains Manager, over which the resource prediction process can be configured. 

The key four main modules of the AI Workload Agent  provide the reasoning controller with distinct methodological approaches to profile the incoming workloads, i.e., 

1) LLM Zero-Shot Estimation: It  utilizes the reasoning \gls{llm} to perform a direct, zero-shot profiling of the AI workload. By analyzing the provided Python script and the characteristics of the linked dataset, the model infers a preliminary forecast of the resource usage. Hereby, it relies solely on its pre-trained knowledge base without any prior execution.

2) Active Profiling (Ground Truth): To establish a reliable ground truth, this module employs an ARIMA forecasting approach based on actual execution metrics. The AI workload, consisting of the script and the data, is packaged into an isolated container and executed within a secured sandbox environment. To ensure a time-efficient execution, it is limited to a subset of the real-world machine learning parameters, such as a reduced number of epochs or a smaller sample size. During this execution, the workload is continuously benchmarked. The gathered historical benchmark data is then fed into the ARIMA model to forecast the resource consumption of the next execution batch. If this prediction converges and falls inside a specific predefined bound, the benchmarking process is halted and the final prediction is calculated. If not, the newly acquired benchmark data is subsequently fed back into the ARIMA model to serve as more data for refining the prediction of the next batch. Finally, the converged ARIMA forecast, together with its corresponding input parameters, is stored persistently in a dedicated profiling knowledge base.

3) LLM + RAG Estimation: It enhances the initial estimation by employing a \gls{rag} mechanism. The stored forecasts from the ARIMA model serve as the foundational RAG database. Using the input JSON, which points to the Python file and the dataset, the \gls{llm} queries this database for similar past executions. By incorporating this historical ground-truth data into the prompt context, the system effectively overcomes the issues of \gls{llm} hallucinations.

4) Re-Profile and Calibrate: The outputs generated by the three previously described workflows are all forwarded to the Constraints Verification agent. This agent checks if the formulated predictions comply with the predefined rules in the Policy and Constraints Manager. If a violation is detected (i.e., the policies and constraints are not ok), this information is routed to the Re-Profile and Calibrate function. This mechanism is capable of re-trigger the orchestrator to reevaluate with additional information. Based on the past executions, the objective hereby is to gather more extensive input data or to establish a new, more accurate ground truth for calibration.

To illustrate the orchestration of these functionalities, an example workflow can be described as the following.
Initially, the policies are configured by the user; for instance, by defining a maximum profiling duration of one minute and setting a nearest-neighbor distance threshold. An incoming AI workload is then sent to the A2A Interface. Subsequently, the reasoning controller analyzes the incoming message and determines the optimal routing to one or more of the available tools.
Assuming the profiling knowledge base is initially empty, the controller is aware that a RAG-based approach will not yield results. Therefore, it forwards the task concurrently to the first tool (LLM Zero-Shot Estimation) and the second tool (Active Profiling). The first tool provides an answer very rapidly, albeit with lower accuracy. Meanwhile, the second tool begins the sandboxed execution but exceeds the configured one-minute time limit. Due to this policy constraint, the answer from the first tool is selected for the immediate prediction, because the second tool is taking too long. However, the second tool continues its operation asynchronously until it finishes the execution, subsequently feeding the gathered ARIMA forecasting data into the profiling knowledge base.
When the next AI workload is posted to the A2A Interface, the reasoning controller performs a k-Nearest Neighbors search within the knowledge base. If the defined distance threshold is met, indicating that a similar workload was already profiled, the task is exclusively routed to the third tool (LLM + RAG Estimation). If the threshold is not met, the controller falls back to utilizing the first and second tools concurrently. As established before, if the second tool fails to finish within the required timeframe, the fast answer from the first tool is used; otherwise, the accurate result from the second tool is preferred. This iterative process continues, thereby continuously enriching the knowledge base and improving the overall prediction accuracy of the framework over time.
\section{System Model and Ground Truth Formulation} 

\begin{figure}
    \centering
    \includegraphics[width=1.0\linewidth]{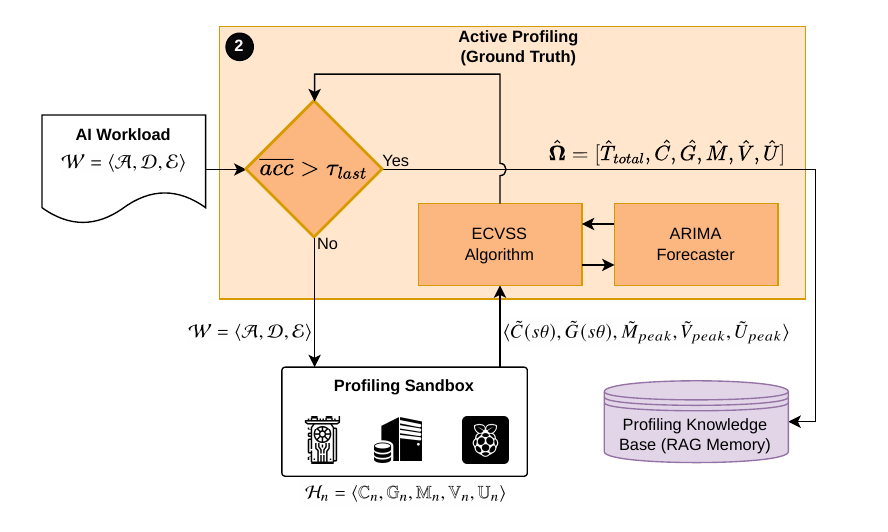}
    \caption{Data flow of the active profiling module and ground truth generation making use of empirical evaluation of AI workloads in a sandboxed environment.}
    \label{fig:ground-truth}
\end{figure}

\begin{table}[htbp]
\centering
\caption{Summary of Key Mathematical Notations}
\label{tab:notation}
\renewcommand{\arraystretch}{1.0} 
\begin{tabularx}{\columnwidth}{@{}l X@{}}
\toprule
\textbf{Symbol} & \textbf{Description} \\
\midrule
\multicolumn{2}{@{}l}{\textit{Hardware \& Resource Constraints}} \\
\midrule
$\mathcal{H}_n$ & Set of resource constraints for edge node $n$ \\
$\mathbb{C}_{n}, \mathbb{G}_{n}$ & CPU capacity (cores/frequencies) and GPU throughput (TFLOPS) \\
$\mathbb{M}_{n}, \mathbb{V}_{n}, \mathbb{U}_{n}$ & System RAM, VRAM, and Total Unified Memory capacity (GB) \\
$c(t), g(t)$ & Instantaneous CPU and GPU utilization at time $t$ \\
$m(t), v(t), u(t)$ & Instantaneous RAM, VRAM, and unified memory utilization at time $t$ \\
$\hat{M}_{peak}, \hat{V}_{peak}, \hat{U}_{peak}$ & Predicted full-scale peak system RAM, VRAM, and Unified Memory \\
$\delta_k(s)$ & Monotonically decreasing memory safety margin for memory pool $k \in \{m, v, u\}$ \\
\midrule
\multicolumn{2}{@{}l}{\textit{Workload \& Profiling Metrics}} \\
\midrule
$\mathcal{W}$ & Incoming AI workload tuple $\langle \mathcal{A}, \mathcal{D}, \mathcal{E} \rangle$ \\
$\mathcal{E}$ & Set of execution environment constraints \\
$N, s$ & Total dataset size and bootstrap sampling ratio $s \in (0, 1]$ \\
$\mathcal{D}_s, B$ & Sub-sampled dataset of size $\lceil s \cdot N \rceil$ and inference batch size \\
$T_{prof}(s)$ & Total profiling overhead time ($T_{init} + T_{load} + T_{exec}$) \\
$T_{init}, T_{load}, T_{exec}$ & Static container initialization, model loading, and dynamic execution time \\
$\tau(x), \bar{\tau}_B$ & Execution latency for sample $x$ and mean batch processing time \\
$\mathbf{R}(t)$ & Multi-variate time-series matrix of resource utilization \\
$\theta, \gamma$ & Telemetry sampling period and number of periods in the bootstrap phase \\
$\Phi$ & Extrapolation function mapping telemetry to full-scale footprint \\
$\hat{\mathbf{\Omega}}$ & Full-scale predicted workload footprint $[\hat{T}_{total}, \hat{C}, \hat{G}, \hat{M}, \hat{V}, \hat{U}]$ \\
\midrule
\multicolumn{2}{@{}l}{\textit{CPU \& GPU Modeling}} \\
\midrule
$C(t), G(t)$ & Instantaneous CPU utilization and GPU occupancy ratio \\
$E$ & Total number of incoming events in the CPU queue \\
$\tilde{C}(s\theta), \tilde{G}(s\theta)$ & Average CPU and GPU utilization over interval $[s\theta, (s+1)\theta]$ \\
$a_e, d_e, p_e$ & Arrival, departure, and processing time of $M/G/\infty$ queue event $e$ \\
$T_{SM}, R_{SM}$ & Maximum resident threads and registers per Streaming Multiprocessor \\
$T_b(t), B_b(t)$ & Number of active resident threads and active blocks per kernel \\
$Z_b(t), R_b(t)$ & Active threads per block and registers required per block \\
$L$ & Total number of observed GPU kernel executions \\
$g_\ell, u_\ell, v_\ell$ & Occupancy level, start time, and completion time of GPU kernel $\ell$ \\
\midrule
\multicolumn{2}{@{}l}{\textit{Forecasting \& Algorithm Constraints}} \\
\midrule
$X_t, Y_t$ & Discrete-time historical sequence and its stationary differenced series \\
$p, d, q$ & ARIMA autoregressive, differencing, and moving average orders \\
$\mathbf{\Phi}, \mathbf{\Theta}$ & Estimated ARIMA parameter vectors for autoregressive and moving average \\
$\mathcal{M}$ & Configuration matrix of dimensions $R \times C$ (epochs $\times$ sample size) \\
$T_{max}$ & Global computational time budget limit for parameter search \\
$\tau_{next}, \tau_{last}$ & Local and global accuracy thresholds for surrogate evaluation \\
$\varrho$ & Target accuracy for parameter search \\
$I$ & Multi-strategy index dictating the traversal leap behavior \\
\bottomrule
\end{tabularx}
\end{table}

\label{sec:system-model}
This section outlines the formal mathematical formulation for the resource consumption of containerized AI workloads on edge nodes prior to full deployment.
Let an incoming AI workload be defined as a tuple $\mathcal{W}$:

\begin{equation}
    \mathcal{W} = \langle \mathcal{A}, \mathcal{D}, \mathcal{E} \rangle, 
\end{equation}
where variable $\mathcal{A}$ represents the algorithm or Python execution graph, while $\mathcal{D}=\{d_k | \forall k\in [1,N] \}$ denotes the complete target dataset (\eg csv, images, text sequences), where $d_k$ is the $k^{\text{th}}$ data point. Finally, $\mathcal{E}=\{\mathcal{E}_{\xi} | \forall \xi \}$ denotes a set of execution environment constraints, i.e. $\mathcal{E}_{1}=1$ if the node supports CUDA, and $0$ otherwise. The \textit{Reasoning Controller} (see \figurename~\ref{fig:Arch} routes the AI workload $\mathcal{W}$ to the \textit{Active Profiling} module depicted in \figurename~\ref{fig:ground-truth} to generate ground truth data.

The bootstrap process creates a sampled execution environment. Let $s \in (0, 1]$ denote the bootstrap sampling ratio.  The samples are randomly chosen, where we define a random mapping $\varphi : [1,s\cdot N] \rightarrow [1,N] | i \rightarrow \varphi(i)$.
We construct a sub-sampled dataset $\mathcal{D}_s=\{ x_{\varphi(1)}, ..., x_{\varphi(s\cdot N)} \} \subset \mathcal{D}$ such that $|\mathcal{D}_s| = \lceil s \cdot N \rceil$. The total profiling overhead time, $T_{prof}(s)$, is composed of a static container initialization overhead ($T_{init}$), model loading time ($T_{load}$), and the dynamic execution time over the sampled dataset ($T_{exec}$):
\begin{equation}\begin{split}
    T_{prof}(s) =&~T_{init} + T_{load} +  T_{exec}, \quad \text{where } \\
    T_{exec} =&~\sum_{i=1}^{\lceil s \cdot N \rceil} \tau(x_{\varphi(i)})
\end{split}
\end{equation}
where $\tau(x_{\varphi(i)})$ is the execution latency for the $i^{\text{th}}$ data sample $x_{\varphi(i)} \in \mathcal{D}_s$. Assuming Independent and Identically Distributed (I.I.D.) samples and a batch size $B$, the dynamic term can be approximated by $\lceil \frac{s \cdot N}{B} \rceil \cdot \bar{\tau}_B$, where $\bar{\tau}_B$ is the mean batch processing time, which will give us an updated execution time, given as.
\begin{equation}\begin{split}
    T_{exec} =& \sum_{i=1}^{\lceil \frac{s \cdot N }{B}\rceil} \tau(x_{\varphi((i-1)B +1)}, ..., x_{\varphi((i-1)B + B)})
\end{split}
\end{equation}

To accurately map this dynamic execution time to physical node saturation, we must formalize the underlying hardware environment. We differentiate between systems with unified memory and systems with discrete memory (system RAM plus VRAM), as edge devices can range from simple Raspberry Pi boards over NVIDIA Jetson systems to standard PCs with a graphics card.  We denote by $\mathcal{H}_n$ the set of resource constraints of node $n$, see \figurename~\ref{fig:ground-truth}.

\begin{equation}
    \mathcal{H}_n = \langle \mathbb{C}_{n}, \mathbb{G}_{n}, \mathbb{M}_{n}, \mathbb{V}_{n}, \mathbb{U}_{n} \rangle
\end{equation}
where $\mathbb{C}_{n}$ is CPU capacity (cores/frequencies) of node $n$, $\mathbb{G}_{n}$ is GPU throughput (TFLOPS), $\mathbb{M}_{n}$ is system RAM in GB, and $\mathbb{V}_{n}$ is VRAM capacity in GB as well. For nodes that share memory, we consider $\mathbb{M}_{n}=\mathbb{V}_{n}=0$, and we define $\mathbb{U}_{n}$ as the total unified memory capacity, which combines both RAM and VRAM of node $n$. 

During the bootstrap execution of the workload $\mathcal{W}$, a telemetry daemon records a multi-variate time-series matrix of resource utilization $\mathbf{R}(t)$ at time $t$, periodically every period $\theta$. We denote by $\gamma= \lfloor \frac{T_{prof}}{\theta}  \rfloor$ the number of periods used in the bootstrap phase. 

\begin{equation}\begin{split}
     \mathbf{R}(t) = [c(t), g(t), m(t), v(t), u(t)]^\top,\\  t \in [\theta, 2\theta,..., \gamma \theta ]
\end{split}
\end{equation}

We define an extrapolation function $\Phi : \mathbb{R}^{5 \times T_{prof}} \times (0, 1] \rightarrow \mathbb{R}^6$ mapping the bootstrap telemetry to the full-scale predicted workload footprint $\hat{\mathbf{\Omega}} = [\hat{T}_{total}, \hat{C}, \hat{G}, \hat{M}, \hat{V}, \hat{U}]$, see \figurename~\ref{fig:ground-truth}. This prediction is the newly generated ground truth data that will be saved in the \gls{rag} database.

Memory consumption in deep learning workloads consists of static allocations (e.g., model weights, container context) and dynamic allocations (e.g., batch activations, gradients). Because the batch size $B$ remains constant between the bootstrap phase and the full-scale execution, the peak memory utilization theoretically stabilizes once the first few batches are processed. 

Consequently, we predict the full-scale peak memory by calculating the at full scale of the observed bootstrap memory footprint. However, to account for continuous memory fragmentation and potential memory leaks across long-running inferences, we introduce a monotonically decreasing safety margin $\delta(s)$.

For architectures with discrete memory pools, the predicted peak system RAM ($\hat{M}_{peak}$) and VRAM ($\hat{V}_{peak}$) are modeled as:

\begin{equation}
    \hat{M}_{peak} = \sup_{t \in [\theta,..., \gamma \theta]} m(t) + \delta_m(s)
\end{equation}
\begin{equation}
    \hat{V}_{peak} = \sup_{t \in [\theta,..., \gamma \theta]} v(t) + \delta_v(s)
\end{equation}
For unified memory architectures, both CPU and GPU processes compete for identical physical pages. The unified memory footprint $u(t)$ captures this shared allocation. The peak unified memory $\hat{U}_{peak}$ is extrapolated as:
\begin{equation}
    \hat{U}_{peak} = \sup_{t \in [\theta,..., \gamma \theta]} u(t) + \delta_u(s)
\end{equation}

For the CPU utilization, similar to \cite{hammer2018queue}, we assume requests that are processed by the CPU are modeled by an $M/G/\infty$ queue. We denote $C(t)$ as the CPU utilization at time point $t$, $a_{1} , ..., a_{E}$ as the arrival times of $E$ incoming events. Similarly,  the departure time is defined as $d_{1} , ..., d_{E}$ with the processing times $p_{1} , ..., p_{E}$ so that $d_{e} = a_{e} + p_{e}$.

\begin{equation}
    \label{math:cpu1}
    C(t)= \sum_{e=1}^{E} I(a_{e} < t < d_{e})
\end{equation}
with $I(X)$ being the indicator function, returning 1 if $X$ is true, otherwise 0.
The average CPU utilization $\tilde{C}(s\theta)$, used in \figurename~\ref{fig:ground-truth} as the input data for the ARIMA forecasting algorithm, can be computed with:

\begin{equation}
    \label{math:cpu2}
    \begin{split}
        \tilde{C}(s\theta) &= \frac{1}{\theta} \int_{s \theta}^{(s+1)\theta} C(t) dt \\
        &= \frac{1}{\theta} \int_{s \theta}^{(s+1)\theta} \sum_{e=1}^{E} I(a_{e} < t < d_{e}) dt \\
        &= \frac{1}{\theta}  \sum_{e=1}^{E} \int_{s \theta}^{(s+1)\theta} I(a_{e} < t < d_{e}) dt \\
        &= \sum_{n=1}^{\theta} [F((s+1)\theta; a_{e}, d_{e}) - F(s\theta; a_{e}, d_{e})]
    \end{split}
\end{equation}
with
\begin{equation}
    F(t; a_{e}, d_{e}) = 
\begin{cases} 
0 & \text{if } t \le a_{e} \\ 
\frac{t - a_{e}}{\theta} & \text{if } a_{e} < t \le d_{e} \\ 
\frac{d_{e} - a_{e}}{\theta} & \text{if } t > d_{e} 
\end{cases}
\end{equation}

We assume that we have received $ \gamma$ observations, $\tilde{C}(\theta), \tilde{C}(2\theta),..., \tilde{C}(\gamma\theta)$. We aim at forecasting the future CPU utilization $\tilde{C}((\gamma+1)\theta), \tilde{C}((\gamma+2)\theta)$,... .

At time $t$, GPU utilization $G(t)$ is a function of GPU card parameters and the resource requirement of the AI workload \cite{pimple2019}. Hence, potential occupancy limitations are imposed by resources such as registers, memory, and the number of streaming multiprocessors (SMs) required by the AI workload. Let $G(t)$ be the instantaneous GPU occupancy ratio, representing the fraction of the GPU thread capacity that is actively occupied, and given by:
\begin{equation}
    \label{math:gpu1}
    \begin{split}
    G(t)= & \frac{\text{Active Thread per Block}}{\text{Thread per SM}} = \frac{T_b(t)}{T_{SM}},\\
    &T_b(t) =  \begin{cases} 
             B_b (t) \cdot Z_{b}(t) & \text{For thread block size } \\
             \frac{R_{SM}}{R_{b}(t)} & \text{For register usage} 
        \end{cases}
    \end{split}
\end{equation}
where $T_{SM}$ is the maximum number of resident threads per SM and $T_b(t)$ is the number of active resident threads induced by the running GPU kernel at time $t$, $B_b(t)$ is the number of active blocks per kernel, $Z_b(t)$ is the number of active threads per block, $R_{SM}$ is the number of registers per SM, $R_b(t)$ is the number of registers required per block by the running kernel. 

Similar to the CPU case, we define the average GPU utilization over the interval $[s\theta,(s+1)\theta]$ as
\begin{equation}
    \label{math:gpu3}
    \tilde{G}(s\theta)
    =
    \frac{1}{\theta}
    \int_{s\theta}^{(s+1)\theta} G(t)dt.
\end{equation}
Assume that $L$ GPU kernel executions are observed within the measurement horizon. Let $u_{\ell}$ and $v_{\ell}$ denote the start and completion times of the $\ell$-th GPU kernel, respectively, and let $g_{\ell}$ denote its corresponding occupancy level. Then, the instantaneous GPU utilization can be written as
\begin{equation}
    \label{math:gpu4}
    G(t)
    =
    \sum_{\ell=1}^{L}
    g_{\ell} I(u_{\ell}<t<v_{\ell}),
\end{equation}
where
\begin{equation}
    \label{math:gpu5}
    g_{\ell}
    =
    \frac{1}{T_{SM}}
    \min
    \left\{
        B_{b,\ell}Z_{b,\ell},
        \frac{R_{SM}}{R_{b,\ell}},
    \right\}.
\end{equation}
Substituting \eqref{math:gpu4} into \eqref{math:gpu3}, the average GPU utilization is obtained as:
\begin{equation}
    \label{math:gpu6}
    \begin{split}
        \tilde{G}(s\theta)
        &=
        \frac{1}{\theta}
        \int_{s\theta}^{(s+1)\theta}
        \sum_{\ell=1}^{L}
        g_{\ell} I(u_{\ell}<t<v_{\ell})dt \\
        &=
        \sum_{\ell=1}^{L}
        g_{\ell}
        \left[
        F_G((s+1)\theta;u_{\ell},v_{\ell})
        -
        F_G(s\theta;u_{\ell},v_{\ell})
        \right],
    \end{split}
\end{equation}
where
\begin{equation}
    \label{math:gpu7}
    F_G(t;u_{\ell},v_{\ell})=
    \begin{cases}
        0, & t \le u_{\ell},\\
        \dfrac{t-u_{\ell}}{\theta}, & u_{\ell}<t\le v_{\ell},\\
        \dfrac{v_{\ell}-u_{\ell}}{\theta}, & t>v_{\ell}.
    \end{cases}
\end{equation}
Therefore, $\tilde{G}(s\theta)$ quantifies the time-normalized GPU occupancy within the $s$-th control interval. We assume that we have received $\gamma$ observations,
$\tilde{G}(\theta),\tilde{G}(2\theta),\ldots,\tilde{G}(\gamma\theta)$,
and we aim at forecasting the future GPU utilization
$\tilde{G}((\gamma+1)\theta),\tilde{G}((\gamma+2)\theta),\ldots$.

\subsection{ARIMA-Based Forecasting Model}
\label{subsec:arima}

Let's denote the discrete-time sequence of historical CPU or GPU utilization measurements as $X_t = \tilde{C}(t\theta)$ or $ \tilde{G}(t\theta)$ for $t \in \{1, 2, \dots, \gamma\}$. To employ the Auto-Regressive Integrated Moving Average (ARIMA) methodology \cite{ibrahimi2017prediction}, we must first stabilize the mean of the time series by applying a differencing operator of order $d$. Let $B$ denote the backshift operator such that $B X_t = X_{t-1}$. The differenced, stationary series $Y_t$ is obtained by:
\begin{equation}
    Y_t = (1 - B)^d X_t
\end{equation}
where $d \in \mathbb{Z}^+$ is the minimum integration order required to achieve stationarity.

Following the stabilization of the series, we model $Y_t$ using an ARMA($p, q$) process, which yields the comprehensive ARIMA($p, d, q$)  ) model for the original sequence $X_t$. This captures the temporal correlation by expressing the current utilization as a linear combination of $p$ past observations (autoregressive components) and $q$ past forecast errors (moving average components), while the integration order $d$ accounts for the number of prior differencing transformations required to render the raw, non-stationary data stationary. The comprehensive ARIMA($p, d, q$) model for CPU utilization is thus formulated as:
\begin{equation}
    \phi_p(B)(1 - B)^d X_t = \theta_q(B) \varepsilon_t
\end{equation}
where $\varepsilon_t \sim \mathcal{WN}(0, \sigma^2)$ is a zero-mean white noise process representing stochastic, unpredictable fluctuations in CPU demand (e.g., microbursts). The autoregressive operator $\phi_p(B)$ and the moving average operator $\theta_q(B)$ are defined as characteristic polynomials of degrees $p$ and $q$, respectively:
\begin{align}
    \phi_p(B) &= 1 - \phi_1 B - \phi_2 B^2 - \dots - \phi_p B^p \\
    \theta_q(B) &= 1 + \theta_1 B + \theta_2 B^2 + \dots + \theta_q B^q
\end{align}

To operationalize this model within the network controller, the structural hyperparameters $(p, d, q)$ must be identified. This is achieved by iteratively evaluating candidate models and minimizing the Akaike Information Criterion (AIC). Once the optimal structure is selected, the parameter vectors $\mathbf{\Phi} = [\phi_1, \dots, \phi_p]^\top$ and $\mathbf{\Theta} = [\theta_1, \dots, \theta_q]^\top$ are estimated using Maximum Likelihood Estimation (MLE) over the historical observation window $[1, \gamma]$.

Finally, to reach a high precision forecast of the resource usage, we compute the $k$-step-ahead forecast of the CPU utilization, denoted as $\hat{X}_{\gamma+k} = \mathbb{E}[X_{\gamma+k} | X_1, \dots, X_\gamma]$. Expanding the generalized difference equation, the one-step-ahead prediction for the imminent time epoch $(\gamma+1)\theta$ is yielded by:
\begin{equation}
    \hat{X}_{\gamma+1} = \sum_{i=1}^{p+d} \alpha_i X_{\gamma+1-i} + \sum_{j=1}^{q} \theta_j \varepsilon_{\gamma+1-j}
\end{equation}
where the coefficients $\alpha_i$ are derived from the algebraic expansion of the combined generalized autoregressive polynomial $\alpha(B) = \phi_p(B)(1-B)^d$. This predictive formulation enables the orchestrator to continuously update $\hat{X}_{\gamma+1}$ as new telemetry data arrives. The same mathematical principles can be applied to the GPU, RAM and VRAM, or uRAM as well.

\subsection{Adaptive Parameter Search Algorithm for Forecasting}
\label{subsec:search}

\begin{algorithm}
\caption{Extended Cross-Validated Surrogate Search}
\label{alg:extended_cv_search}
\KwIn{
    Matrix $\mathcal{M}$ of size $R \times C$, global limit $T_{max}$,\\
    \quad\quad Local threshold $\tau_{next}$, Global threshold $\tau_{last}$,\\
    \quad\quad Split ratio $\gamma \in (0, 0.8]$, Strategy Index $I \in \{1, 2, 3\}$
}
\KwOut{Final estimated resource $\hat{\Omega}_{R,C}$}
\BlankLine

$H \leftarrow \emptyset$; $t_{total} \leftarrow 0$; $c_{skip} \leftarrow 0$; $F_{future} \leftarrow \emptyset$\;
\BlankLine

\For{$c \leftarrow 0$ \KwTo $C - 1$}{
    \If{$c_{skip} > 0$}{
        $c_{skip} \leftarrow c_{skip} - 1$; \textbf{continue}\;
    }
    
    $r_{skip} \leftarrow 0$\;
    
    \For{$r \leftarrow 0$ \KwTo $R - 1$}{
        \If{$r_{skip} > 0$}{
            $r_{skip} \leftarrow r_{skip} - 1$; \textbf{continue}\;
        }
        
        $\Omega_{r,c}, t_{exe} \leftarrow \textsc{ComputeResources}(\mathcal{M}[r, c])$\;
        $H \leftarrow H \cup \{\Omega_{r,c}\}$\;
        $t_{total} \leftarrow t_{total} + t_{exe}$\;
        \If{$t_{total} \ge T_{max}$}{ \Return{\texttt{TIMEOUT}, $F_{future}[|F_{future}|-1].\hat{\Omega}$} }
        
        \uIf{$|H| < 8$}{
            \textbf{continue} \tcp*{Mandatory Baseline}
        }
        \Else{
            $status, \hat{\Omega}_{res}, L_c, L_r, F_{future} \leftarrow \textsc{EvaluateSurrogateLeap}$
            $(H, \mathcal{M}, c, r, \Omega_{r,c}, \gamma, \tau_{last}, \tau_{next}, I, R)$\;
            
            \If{$status = \texttt{SUCCESS}$}{
                \Return{\texttt{SUCCESS}, $\hat{\Omega}_{res}$}\;
            }
            \If{$status = \texttt{BREAK\_COL}$}{
                $c_{skip} \leftarrow L_c$; \textbf{break}\;
            }
            \If{$status = \texttt{LEAP\_ROW}$}{
                $r_{skip} \leftarrow L_r$\;
            }
        }
    }
}
\BlankLine\Return{\texttt{EXHAUSTED}, $F_{future}[|F_{future}|-1].\hat{\Omega}$}\;
\end{algorithm}

\begin{algorithm}
\caption{Evaluate Surrogate and Leap Strategy}
\label{alg:eval_surrogate}
\KwIn{$H, \mathcal{M}, c, r, \Omega_{r,c}, \gamma, \tau_{last}, \tau_{next}, I, R$}
\KwOut{Tuple $(status, \hat{\Omega}_{res}, L_c, L_r, F_{future})$}
\BlankLine

\tcp{Define Data Partitions}
$v_{idx} \leftarrow \lfloor \gamma \times |H| \rfloor$\;
$H_{train} \leftarrow H[0 : v_{idx}-1]$\;
$H_{val} \leftarrow H[v_{idx} : |H|-1]$\;
$U \leftarrow \{(x,y) \in \mathcal{M} \mid y > c \lor (y = c \land x > r)\}$\;

$F_{val}, F_{future} \leftarrow \textsc{ComputeARIMA}(H_{train}, H_{val}, U)$\;

\tcp{With $(acc_{i}, \hat{\Omega}_{i}) \leftarrow F[i]$}
$acc_{next} \leftarrow F_{val}[0].acc$\;
$\overline{acc} \leftarrow \frac{1}{|F_{val}|} \sum_{i=0}^{|F_{val}|-1} F_{val}[i].acc$\;

\BlankLine
\If{$\overline{acc} > \tau_{last}$}{ 
    \eIf{$U \neq \emptyset$}{
        \Return{(\texttt{SUCCESS}, $F_{future}[|F_{future}|-1].\hat{\Omega}, 0, 0, F_{future}$)}\;
    }{
        \Return{(\texttt{SUCCESS}, $\Omega_{r,c}, 0, 0, F_{future}$)} \tcp*{Matrix exhausted}
    }
}

\BlankLine
\If{$acc_{next} > \tau_{next}$}{
    $skip\_col, L_c, L_r \leftarrow \textsc{CalcLeapDistance}(c, r, I, R)$\;
    \eIf{$skip\_col$}{
        \Return{(\texttt{BREAK\_COL}, \texttt{NULL}, $L_c, 0, F_{future}$)}\;
    }{
        \Return{(\texttt{LEAP\_ROW}, \texttt{NULL}, $0, L_r, F_{future}$)}\;
    }
}

\BlankLine
\Return{(\texttt{CONTINUE}, \texttt{NULL}, $0, 0, F_{future}$)}\;
\end{algorithm}

\begin{algorithm}
\caption{Calculate Multi-Strategy Leap}
\label{alg:calc_leap_indexed}
\KwIn{Current col $c$, Current row $r$, Strategy Index $I$, Total rows $R$}
\KwOut{Tuple $(skip\_col, L_c, L_r)$ representing column abort, column leap, and row leap}
\BlankLine

\uIf{$I = 1$}{
    \tcp{Strategy 1: No leaping, exhaustive search}
    \Return{(\textbf{false}, 0, 0)}\;
}
\uElseIf{$I = 2$}{
    \tcp{Strategy 2: Intra-column row leaping}
    \Return{(\textbf{false}, 0, 3)}\;
}
\ElseIf{$I = 3$}{
    \tcp{Strategy 3: Inter-column tiered leaping}
    $d \leftarrow (r + 1) / R$ \tcp*{Calculate depth percentage}
    
    \uIf{$d \le \frac{1}{3}$}{ \Return{(\textbf{true}, 3, 0)} }
    \uElseIf{$d \le \frac{2}{3}$}{ \Return{(\textbf{true}, 2, 0)} }
    \Else{ \Return{(\textbf{true}, 1, 0)} }
}
\BlankLine
\Return{(\textbf{false}, 0, 0)} \tcp*{Fallback safety condition}
\end{algorithm}

To find the best possible parameter combination with which the ARIMA-based forecasting model is called, we implement an adaptive parameter search over the size of the training data and the number of epochs that the model will be trained on during the profiling phase, as depicted by the decision structure on the left of the \textit{Active Profiling} module in \figurename~\ref{fig:ground-truth}.
These parameter combinations are denoted in matrix $\mathcal{M}$, where the columns represent the training samples $n_{samples}$ in steps of 1000 samples, and the rows represent the number of epochs $n_{epochs}$, always incremented by one. The start point $(0,0)$ of the matrix resembles 1 epoch and 1000 training samples.
The forecasting model is than initialized with $n_{samples}$ and $n_{epochs}$ to predict the future resource usage $\hat{\Omega}$, but will as well return an accuracy score about the prediction, which we will use in the following to calculate the steps through the matrix $\mathcal{M}$ to find the best possible combination for a efficient forecast, while respecting the time constrains $T_{max}$ of the request, to keep the forecasting in a reasonable time frame.
To efficiently locate regions meeting the accuracy requirement within the matrix $\mathcal{M}$, we propose the Extended Cross-Validated Surrogate Search (ECVSS) (summarized in Algorithm \ref{alg:extended_cv_search}). Unlike traditional exhaustive search methods that evaluate continuous sub-optimal regions, the proposed approach introduces a dynamic forward-leaping mechanism. By mathematically bypassing cells that can be predicted with a high accuracy, the algorithm rapidly converges on the target accuracy $\varrho$ while strictly adhering to a computational time budget $T_{max}$.

To achieve this, ECVSS traverses $\mathcal{M}$ in a column-major sequence using an active learning loop. At its core, the algorithm operates on a ``Compute-First'' principle. For any evaluated spatial coordinate $(r, c)$, the system mandatorily executes the exact resource computation to yield the ground-truth state $\Omega_{r,c}$, appending this value to a historical time-series vector $H$. To prevent predictive instability caused by a cold-start, the initial $n_{samples}$ coordinates are strictly computed without querying the forecasting engine, thereby establishing a robust empirical baseline. 

Once the empirical baseline is established, ECVSS dynamically transitions into a rolling cross-validation paradigm (detailed in Algorithm \ref{alg:eval_surrogate}). At each sequential coordinate step, the accumulated historical array $H$ is partitioned into a training set $H_{train}$ and a validation set $H_{val}$ using a proportional split ratio $\gamma$. To mathematically guarantee the existence of validation data and prevent fatal division-by-zero anomalies during evaluation, this ratio is strictly bounded to the interval $\gamma \in (0, 0.8]$. Concurrently, the algorithm constructs a strictly forward-looking uncomputed queue $U \leftarrow \{(x,y) \in \mathcal{M} \mid y > c \lor (y = c \land x > r)\}$. The surrogate \textsc{ARIMA} model is trained exclusively on $H_{train}$ to generate two distinct outputs: a scored predictive array over the validation set ($F_{val}$), which yields both estimated resources and empirical accuracy metrics, and an unscored macroscopic projection over the entire uncomputed future space ($F_{future}$).

By decoupling the scoring domain from the predictive domain, the algorithm rigorously evaluates model reliability on known ground-truth data without sacrificing its capacity for distant projection. The system calculates the mean accuracy of the validation forecasts, $\overline{acc}$. As shown in \figurename~\ref{fig:ground-truth}, if $\overline{acc}$ exceeds the strict global threshold $\tau_{last}$, it is assumed that the next predictions are done very accurately. At this point, the algorithm has enough confidence in the accuracy of the estimate for the resource usage to execute an early global termination. To eliminate predictive redundancy, the algorithm implements a semantic safeguard: if the matrix is fully exhausted ($U = \emptyset$), it yields the exact ground-truth state $\Omega_{r,c}$ with a \texttt{SUCCESS} flag. Otherwise, it projects to the absolute end of the uncomputed space, returning the final estimated resource $F_{future}[|F_{future}|-1].\hat{\Omega}$, successfully resolving the search while bypassing all remaining computations.

If global convergence is not achieved, the subroutine evaluates the localized confidence of the immediate next validation point, defined as $acc_{next}$, to return a state control flag to the main traversal loop. An uncertain local forecast ($acc_{next} \le \tau_{next}$) yields a \texttt{CONTINUE} flag, forcing the main algorithm to advance to the subsequent adjacent coordinate and perform an exact computation to iteratively reinforce $H_{train}$ with new data. Conversely, a confident localized forecast ($acc_{next} > \tau_{next}$) queries the leaping logic (see Algorithm \ref{alg:calc_leap_indexed}). We compare the performance of different leaping strategies. The magnitude and dimensional axis of this leap are dictated by a modular Strategy Index $I$. Depending on the selected index, the heuristic can return \texttt{CONTINUE} to enforce exhaustive verification ($I=1$), return \texttt{LEAP\_ROW} to execute an intra-column vertical leap to rapidly descend through the epoch parameter space ($I=2$), or return \texttt{BREAK\_COL} to perform an aggressive inter-column horizontal leap ($I=3$). For $I=3$, the horizontal magnitude is inversely proportional to the fractional matrix depth $d = (r+1)/R$, penalizing discoveries deep within the epoch rows to heavily incentivize early-column traversal.

Throughout the active search loop, the cumulative execution time $t_{total}$ is continuously evaluated against the global constraint $T_{max}$. If the computational budget is exceeded, the algorithm triggers a \texttt{TIMEOUT} interrupt. To ensure graceful degradation, the algorithm avoids catastrophic failure by returning the furthest available projected resource state $F_{future}[|F_{future}|-1].\hat{\Omega}$. Finally, if the traversal loop mathematically completes the entire configuration space without triggering global convergence or a timeout, the system outputs an \texttt{EXHAUSTED} flag alongside the final projection, ensuring strict semantic differentiation between a time-constrained interrupt and a fully traversed configuration search.
\section{Data}
\label{sec:data}
To create the \gls{ai} training data, we created 8 \gls{ai} workloads that train on 6 different well known datasets. For the sake of comparative analysis, we distinguish between visual and tabular data and opt for classification algorithms only. The workloads ran on gls{rpi}, Thor Jetson and workstation PC, as examples of data for three different hardware options. The training was carried out with an increasing sample size and ran for 20 or 30 epochs, depending on the hardware. We chose the values with the highest epoch for the results shown. As can be seen in Table \ref{tab:dataset_specs} different datasets contain different amounts of data. Therefore, we chose to cap the Vision datasets to 30000 samples. The tabular datasets were capped at different lengths: The HIGGS dataset at 30000 rows, the forest coverage dataset at 10000 rows and the wine quality dataset at 4900 rows. This was necessary to prevent overfitting. The  metrics measured include training time, testing time as well as GPU-, CPU-, RAM- and VRAM- or uRAM usage. While the training and testing time can be easily logged, for the GPU, CPU and different memory, we sampled the usage values every 10ms, which we estimate does not not put additional stress on the CPU. The data generated is then used to calculate the mean usage as well as the peak usage for the three metrics and is written in the dataset. This dataset also contains the measured testing accuracy as well as the testing loss. In the following we show some of the results from our least powerful and our most powerful device which also validate our approach.
The full datasets are openly available and can be found \cite{marlagruedgedevicedataset}.

\subsection{RPI}
\begin{figure}
    \centering
    \includegraphics[width=1\linewidth]{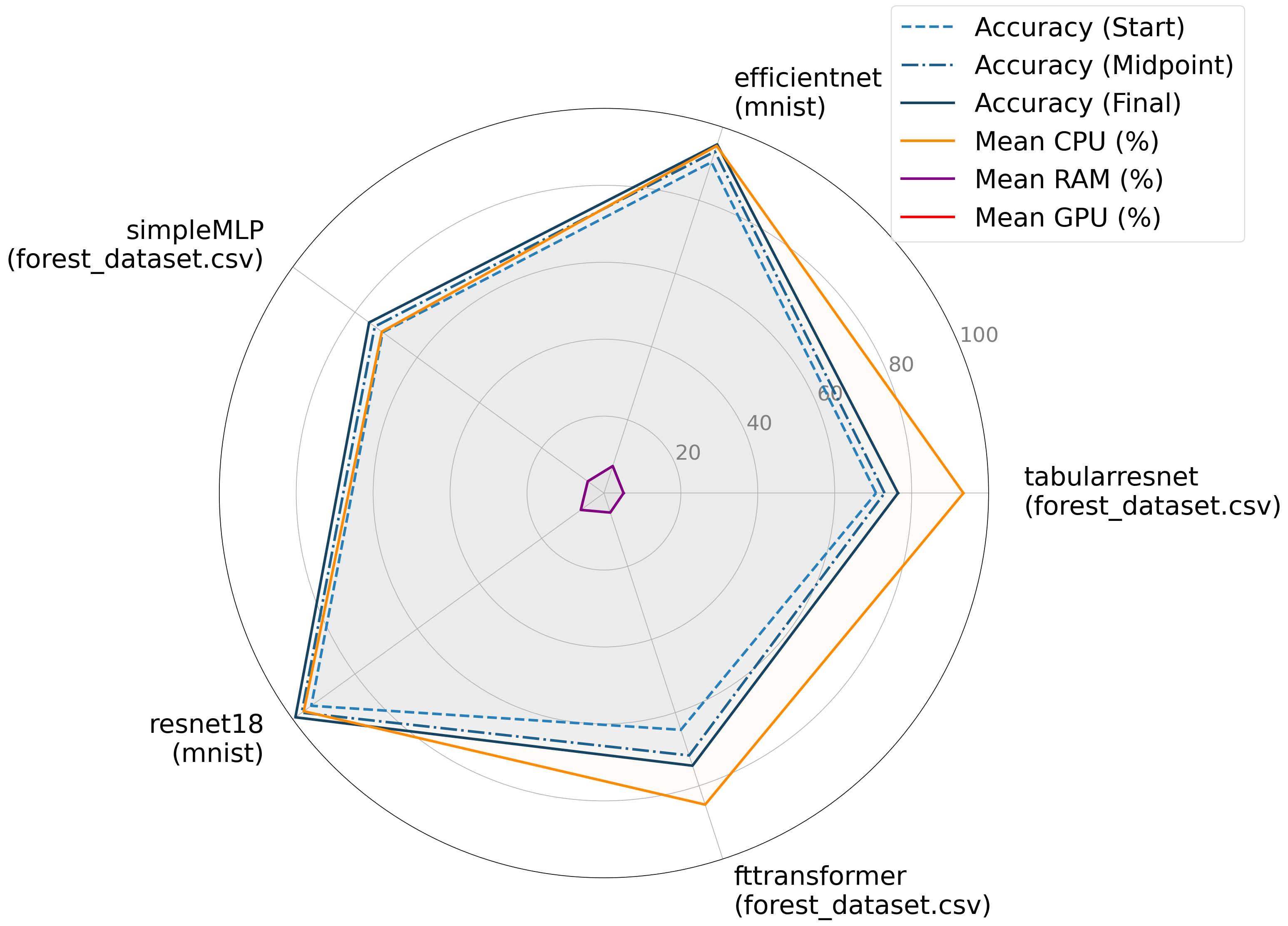}
    \caption{Accuracy, CPU usage and RAM usage for model training on the \gls{rpi}.  The combinations of model and dataset are shown on the outside, with the model name above and the dataset name in brackets. Tabular workloads can be distinguished by the addition of the CSV file type.  As the device has no GPU, no GPU performance is shown.  }
    \label{fig:RPI_Spider}
\end{figure}

\begin{figure}
    \centering
    \includegraphics[width=1\linewidth]{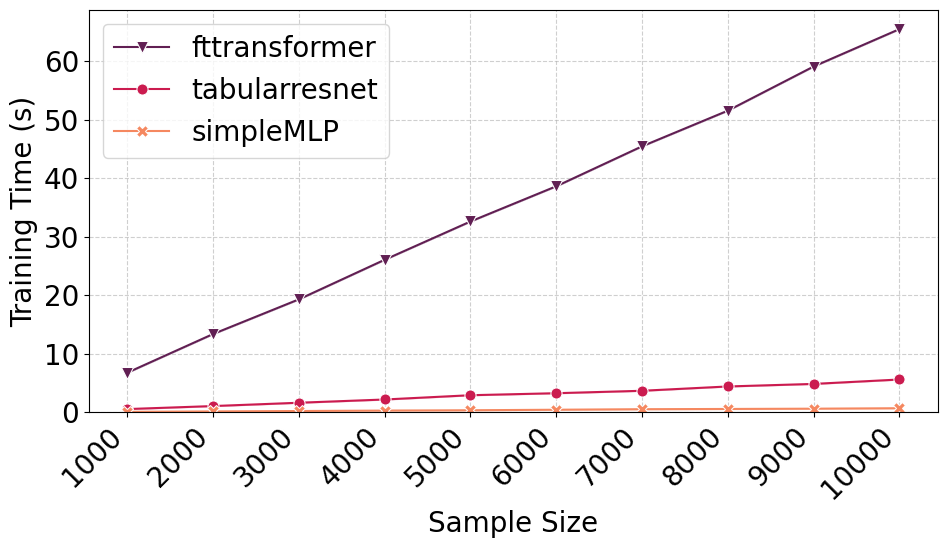}
    \caption{Training time per epoch over sample size for the tabular workload, for each of the tested models on the forest dataset, for 20 epochs on the \gls{rpi}.  }
    \label{fig:RPI_Forest}
\end{figure}

\begin{figure}
    \centering
    \includegraphics[width=1\linewidth]{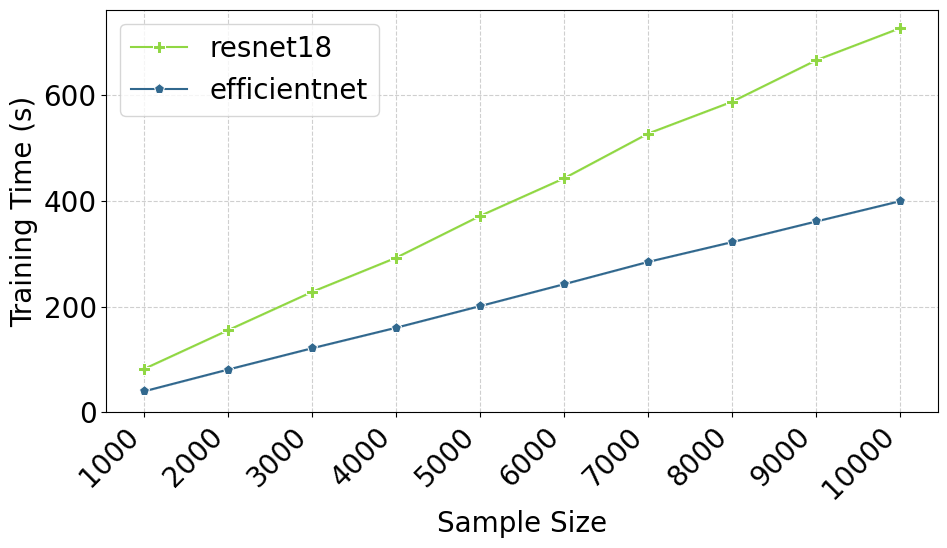}
    \caption{Training time per epoch over sample size for the vision workload, for each of the tested models on the MNIST dataset, for 20 epochs on the \gls{rpi}.}
    \label{fig:RPI_MNIST}
\end{figure}

Dataset generation was conducted using a \gls{rpi} 5 equipped with 16 GB of RAM, representing the highest computational capacity currently available in this hardware series. Despite these specifications, the device remains resource-constrained for large-scale AI training. Consequently, the experimental scope was restricted to lightweight models with minimal parameter counts and limited to two datasets to mitigate onboard storage limitations. Furthermore, model training was capped at a maximum of 20 epochs, as extended processing times rendered the system impractical for real-world deployment.
\figurename~\ref{fig:RPI_Spider} shows the testing accuracy of the first and last epoch as well as the middle epoch used, as well as the mean CPU, RAM, GPU and VRAM. Since the \gls{rpi} does not have a  GPU, the values for GPU and VRAM usage are always zero. What can be observed from this image, is that the main training bottleneck on the \gls{rpi} is the CPU. 
Training vision AI workloads on CPU only, is possible, but time consuming as can be seen in Figures \ref{fig:RPI_Forest} and \ref{fig:RPI_MNIST} which shows the training time per epoch over the increasing sample size, each for the maximum epochs of 20. The figures show the dependencies between the visual and tabular models, with the highest training time per epoch reaching over 10 minutes for ResNet18 and the lowest being less than one second for simpleMLP.

Overall, \figurename~\ref{fig:RPI_Spider} to \figurename~\ref{fig:RPI_MNIST} show that the \gls{rpi} is capable of training vision and tabular data feed for \gls{ai} models. However, the training process is very time and CPU consuming especially for the vision tasks. 


\subsection{JETSON}
\begin{figure*}[h]
    \centering
    \includegraphics[width=0.9\linewidth]{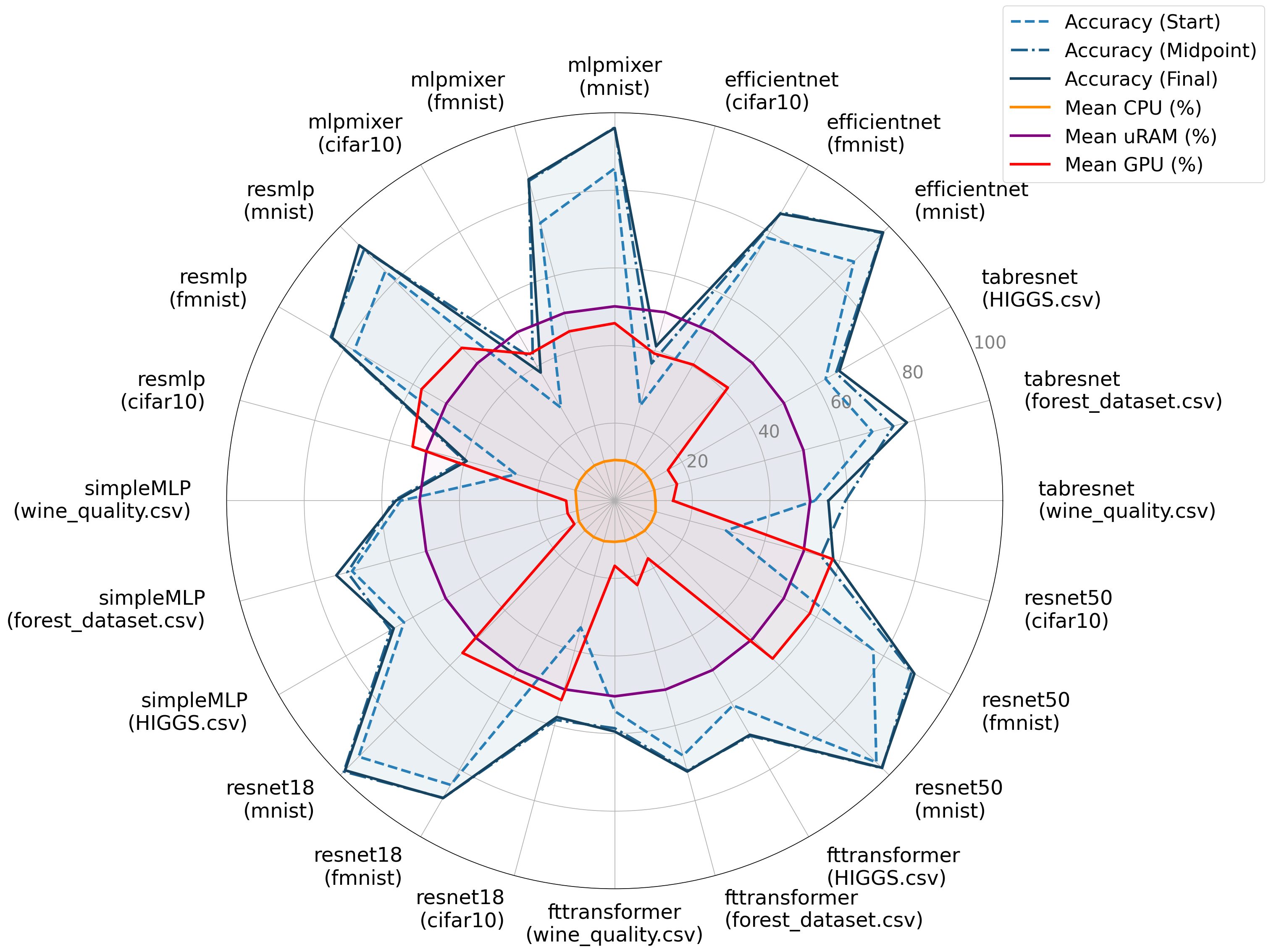}
    \caption{Testing Accuracy, Mean CPU, GPU and uRAM usage for the tested visual and tabular workloads on the JETSON Thor. The combinations of model and dataset are shown on the outside, with the model name above and the dataset name in brackets. Tabular workloads can be distinguished by the addition of the CSV file type. The uRAM and CPU values are constant for all combination, since training is carried out using the JETSONs GPU.}
    \label{fig:JEFFSpider}
\end{figure*}

\begin{figure}
    \centering
    \includegraphics[width=1\linewidth]{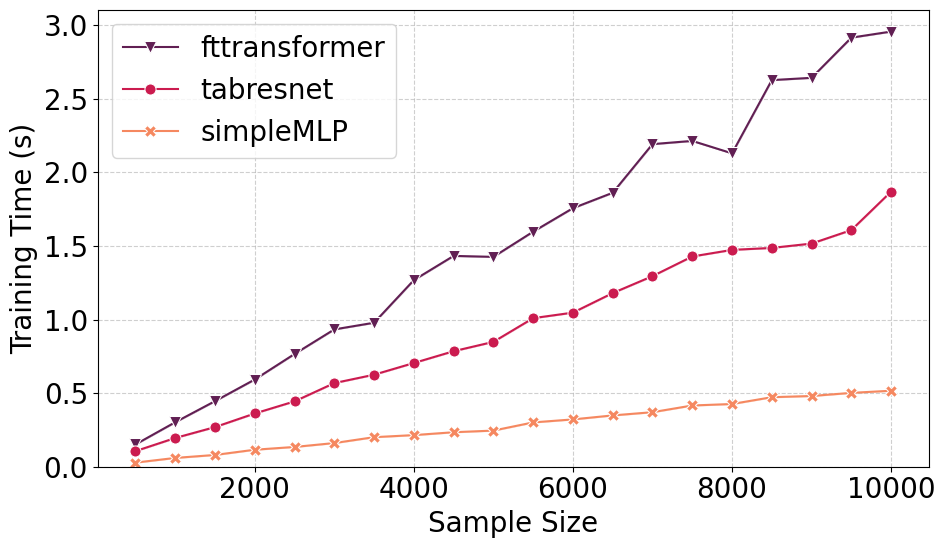}
    \caption{Training time per epoch over sample size for the forest dataset on the JETSON Thor, running for 30 epochs. }
    \label{fig:JEFF_Forest}
\end{figure}

\begin{figure}
    \centering
    \includegraphics[width=1\linewidth]{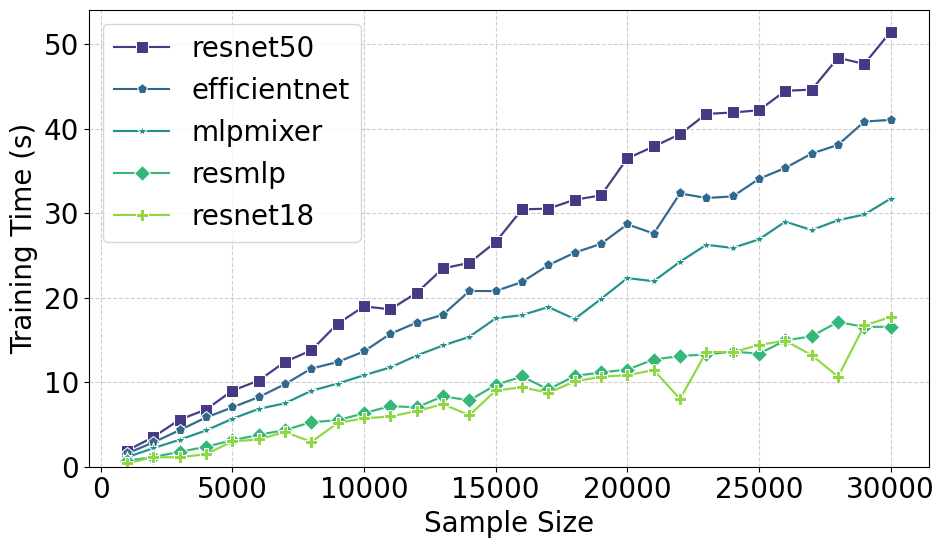}
    \caption{Training time per epoch over sample size for the MNIST dataset on the JETSON Thor, running for 30 epochs.}
    \label{fig:JEFF_MNIST}
\end{figure}

\begin{figure}
    \centering
    \includegraphics[width=1\linewidth]{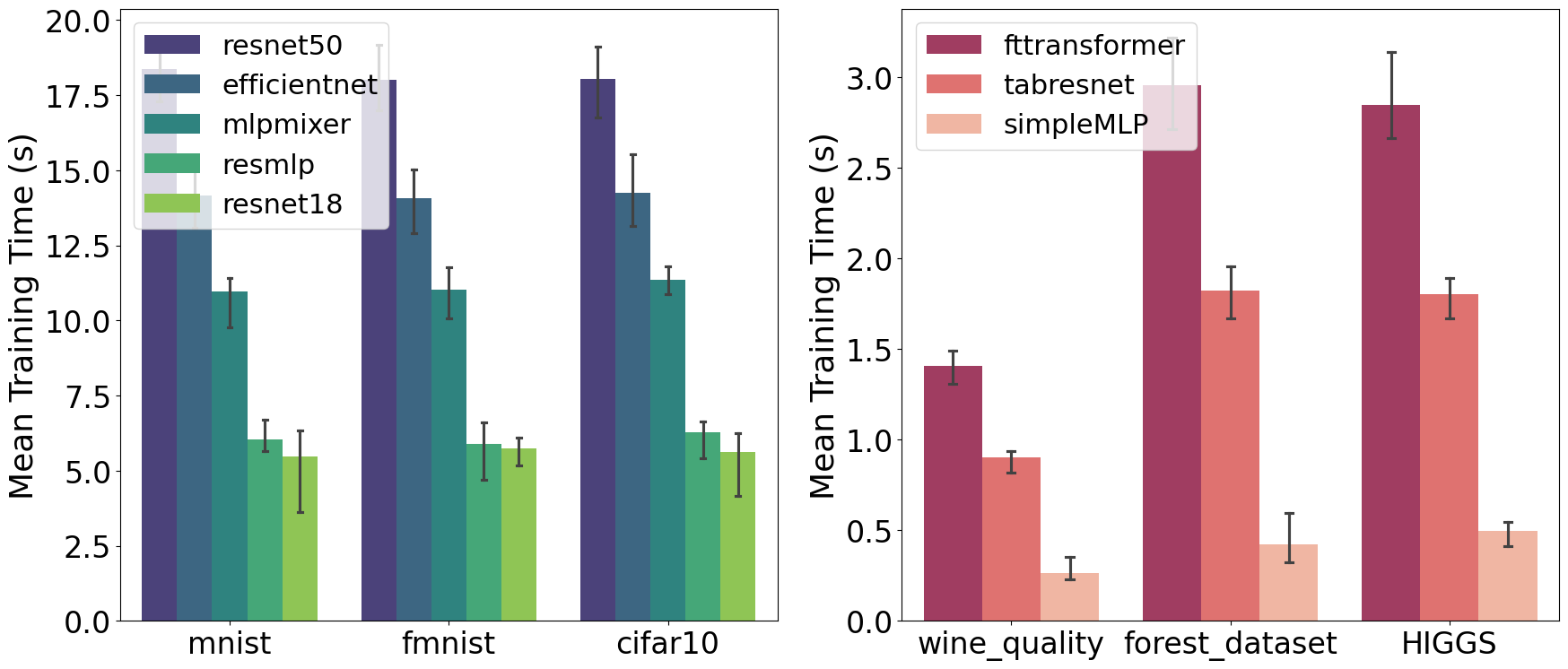}
    \caption{Mean training times per epoch for the visual and tabular datasets on the JETSON, showcasing all data and model combinations. The choice of data does not have a major influence on the mean training time per epoch. }
    \label{fig:JEFF_MeanTrainingTime}
\end{figure}

NVIDIA Jetson Thor, equipped with an NVIDIA Blackwell architecture GPU and 128 GB of unified memory (uRAM), has an enhanced computational capacity for experimentation, encompassing eight distinct models evaluated across three datasets each, as detailed in Tables \ref{tab:dataset_specs} and \ref{tab:algorithm_specs_short}.
For these experiments, training durations were extended to a maximum of 30 epochs. Figure \ref{fig:JEFFSpider} illustrates the accuracy achieved at the initial, middle, and final epochs for all model-dataset configurations, alongside their corresponding mean GPU-, CPU-, and uRAM utilization.
The results underscore the importance of optimizing training parameters; certain architectures exhibited near-optimal performance after a single epoch and showed marginal gains with prolonged training. For instance, MLP-Mixer demonstrated rapid convergence on MNIST dataset, and similarly showed no significant performance disparity between 15 and 30 epochs on CIFAR-10 dataset. Furthermore, the empirical data highlights divergence in GPU utilizations: vision-based models required two to three times more of the GPU resources than tabular models, whereas CPU and uRAM utilization remained highly consistent in all experimental combinations. 
The observed uRAM usage presents the entire observed system usage and not only the neural network footprint. By refining our measurement methodology to isolate process-specific memory and PyTorch's internal allocator, we successfully bypassed these system-level artifacts and observed that AI workloads allocated between 0.8\% to 2.0\% of the Jetson Thor's uRAM. Our data reflects the actual hardware behavior, which allocates 50\% uRAM accurately describing the true operational baseline.
Because all training and inference computations were explicitly delegated to  GPU,  CPU was primarily relegated to data loading and parameter management.
\figurename~\ref{fig:JEFF_Forest} and \figurename~\ref{fig:JEFF_MNIST} show the training time per epoch as the sample size increases on the Jetson platform for the eight evaluated tabular (Forest dataset) and visual (MNIST dataset) workloads. In Figure \ref{fig:JEFF_Forest}, the sample size is capped to 10000, which is the total cardinality of the dataset. Increasing this limit through oversampling would lead to severe model overfitting. Comparing these results with those presented in \figurename~\ref{fig:RPI_Forest} and \figurename~\ref{fig:RPI_MNIST} shows that the Jetson provides substantial computational acceleration, processing visual and tabular workloads 10 and 20 times faster than the previously tested hardware, respectively. Finally, Figure \ref{fig:JEFF_MeanTrainingTime} summarizes the mean per-epoch training latency across all evaluated model-dataset combinations at a fixed sample size of 10000. For the visual dataset, it shows that the choice of dataset has very little impact on mean training time. The same is true for the tabular data, with the wine quality dataset being the only outlier. This is because it is the smallest dataset, with the average being calculated for 5000 samples.

\section{Evaluation} 
\label{sec:eval}
This section describes the experimental evaluation of the AI Workload Agent system, comparing the full agentic architecture against a zero-shot LLM baseline.
\subsection{Experimental Setup}
\label{subsec:experimental-setup}
\begin{figure}[htbp]
    \centering
    
    \begin{subfigure}[b]{0.24\textwidth}
        \centering
        \includegraphics[width=\textwidth]{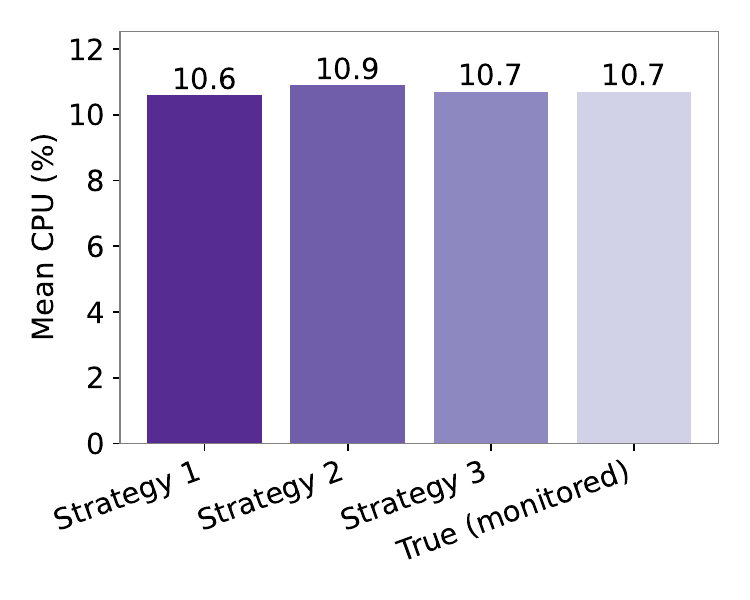}
        \caption{Forecasted vs actual CPU values}
        \label{fig:arima_cpu}
    \end{subfigure}%
    \hfill
    \begin{subfigure}[b]{0.24\textwidth}
        \centering
        \includegraphics[width=\textwidth]{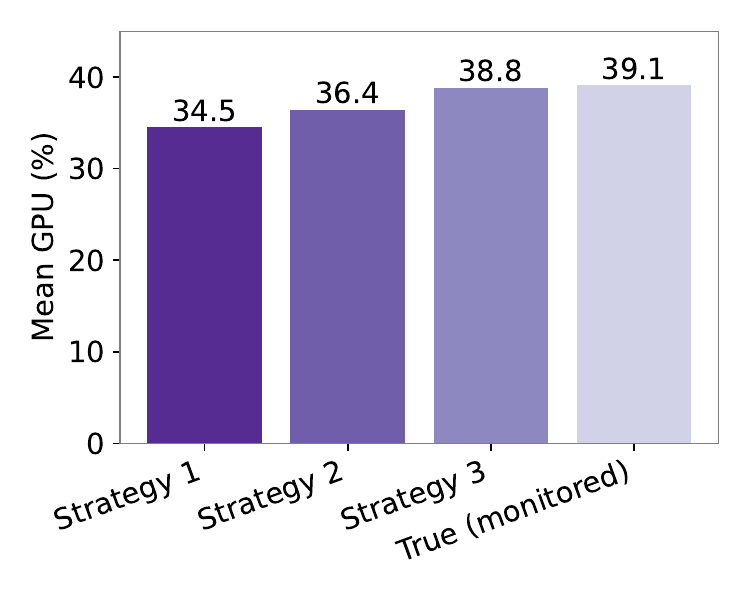}
        \caption{Forecasted vs actual GPU values}
        \label{fig:arima_gpu}
    \end{subfigure}
    
    \vspace{1em}
    
    \begin{subfigure}[b]{0.24\textwidth}
        \centering
        \includegraphics[width=\textwidth]{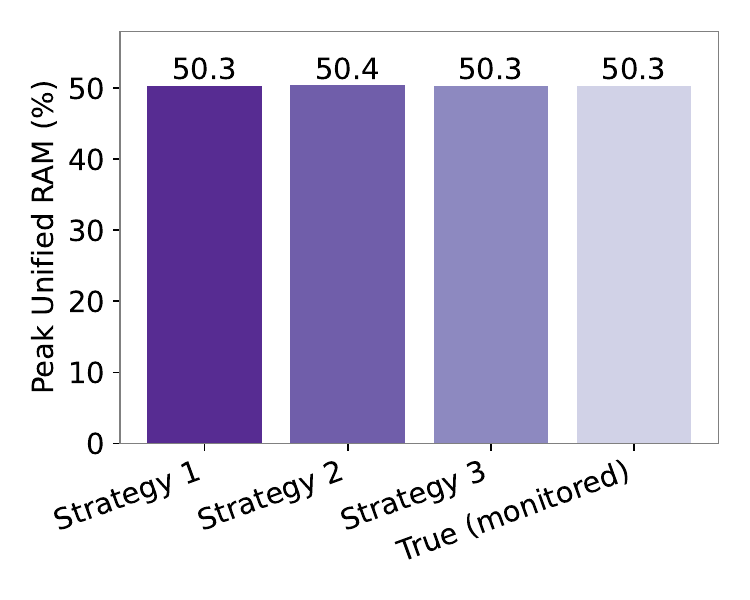}
        \caption{Forecasted vs actual uRAM values}
        \label{fig:arima_ram}
    \end{subfigure}%
    \hfill
    \begin{subfigure}[b]{0.24\textwidth}
        \centering
        \includegraphics[width=\textwidth]{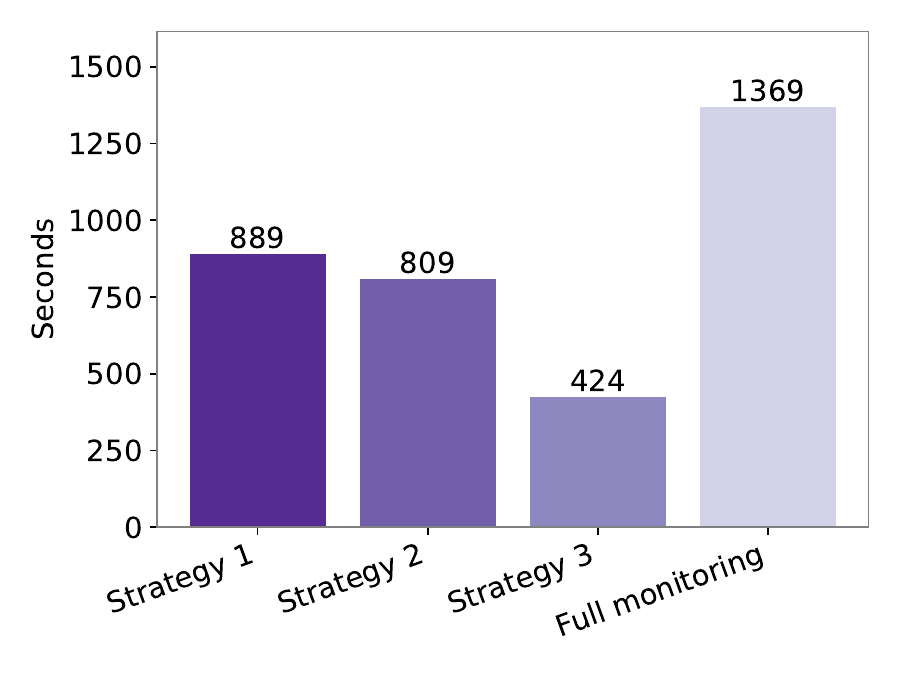}
        \caption{Time to successful forecast, full monitoring vs ARIMA}
        \label{fig:arima_time}
    \end{subfigure}   
    
    \caption{Corresponding ARIMA values for CPU, GPU, uRAM, and Execution Time for EfficientNet, cifar10 on the Jetson Thor.}
    \label{fig:arima_all_metrics}
\end{figure}

Two prediction approaches are evaluated and compared against ground truth data:
\textit{1.) Zero-Shot LLM Baseline.} A simplified pipeline that routes all workloads exclusively through the LLM Zero-Shot Estimation node, bypassing the orchestrator, k-NN search, Active Profiling, and RAG components. This establishes a lower-bound baseline representing pure LLM inference without any system-level augmentation.
\textit{2.) Full Agentic Architecture.} The complete pipeline as described in Section~\ref{sec:arch}, comprising: A2A Interface; Policy Manager; k-NN Similarity Search on the profiling knowledge base; Reasoning Orchestrator (which routes to LLM Zero-Shot, Active Profiling, or LLM+RAG based on k-NN match confidence); Constraints Verification; and Re-Profile and Calibrate on policy violations.
All predictions are evaluated against pre-computed ARIMA telemetry data containing 53 entries with measured values for training time, CPU usage, GPU usage, RAM usage and VRAM usage or unified RAM usage. 
All ARIMA strategies post an improvement over the full monitoring method, as the full monitoring would results in much higher latencies for the full assessment of the hardware values. With the ARIMA strategies we reach very comparable forecasts to the real world resource usage (see \figurename~\ref{fig:arima_cpu}, \ref{fig:arima_gpu} and \ref{fig:arima_ram}), but while the true monitoring time of the training task for the Cifar10 dataset to train an EfficientNet on a Jetson Thor would take 1400 seconds the forecast times are between 425 to 890 seconds (see \figurename~\ref{fig:arima_time}). We proved that it achieves comparable performance to Strategy 1 and the real world telemetry data, but with significantly less time. Averaged over all ARIMA forecasts Strategy 1 needs 693.22 seconds, Strategy 2 needs 473.61 seconds and Strategy 3 only takes 166.97 seconds. So for the ARIMA forecasting in the Full Agentic Architecture we are only using strategy 3.

All LLM inference is performed using the \texttt{mistralai/Magistral-Small-2509} model. This is a 7B parameter model optimized for instruction following and tool-use, making it suitable for the agentic pipeline's structured JSON output requirements.
\subsection{Parameter Configuration}
\label{subsec:parameter-configuration}
Table~\ref{tab:parameter-config} lists all tunable parameters with their values and justifications. These settings are derived from the system's configuration files and agent implementations.

\begin{table*}[htbp]
  \centering
  \caption{AI Workload Agent Parameter Configuration}
  \label{tab:parameter-config}
  \begin{tabular}{@{}llp{11.5cm}@{}}
  \toprule
  \textbf{Category} & \textbf{Parameter} & \textbf{Value / Justification} \\
  \midrule
  \multirow{3}{*}{LLM}
  & Model & \texttt{mistralai/Magistral-Small-2509} (7B parameters, selected for instruction following and tool-use capability) \\
  & Provider & TU Braunschweig KI-Toolbox API (local deployment, ensuring reproducibility and data privacy) \\
  & Max Retries & 3 (schema validation attempts; balances recovery from format errors against API latency) \\
\midrule
\multirow{4}{*}{RAG}
  & Embedding model & \texttt{BAAI/bge-m3} (multilingual dense retrieval, selected for its strong performance on technical text similarity) \\
  & Vector store & ChromaDB (in-memory, lightweight, sufficient for the growing KB of $<$100 entries) \\
  & Chunk size / overlap & 1000 / 200 (standard settings for JSON-structured telemetry documents, preserving field boundaries) \\
  & Top-$k$ retrieval & 3 (limits retrieved context to the most relevant historical profiles, reducing token consumption) \\
\midrule
\multirow{6}{*}{k-NN Search}
  & $k$ (neighbors) & 3 (small value appropriate for a knowledge base that grows from 0 to 53 entries during the experiment) \\
  & Distance metric & Manhattan (L1 norm, robust to categorical feature mismatches and scale-invariant after preprocessing) \\
  & Threshold mode & \texttt{closest} (only the nearest neighbor's distance is checked against the threshold, most conservative match) \\
  & Distance threshold & 2.0 (empirically tuned; balances false positives (overly broad matches) against false negatives (missing valid RAG candidates)) \\
  & Min matches required & 2 (ensures at least two neighbors exist before declaring a threshold match, preventing single-outlier reliance) \\
  & Categorical features & Algorithm, Dataset, Device (one-hot encoded) \\
  & Numerical features & Samples, Epochs) \\
\midrule
\multirow{3}{*}{Policy}
  & Max profiling duration & 60~s (hard upper bound for Active Profiling telemetry retrieval; triggers timeout fallback to Zero-Shot) \\
  & Max total forecasting time & 600~s (absolute ceiling for the entire prediction pipeline, including retries and re-profiling) \\

\midrule
\multirow{2}{*}{Re-Profile}
  & Max re-profile iterations & 3 (prevents infinite loops on persistent constraint violations; after 3 failures, the pipeline returns the best-effort prediction) \\
\midrule
\multirow{3}{*}{Batch}
  & Workloads & 53 (complete set, covering all algorithm-dataset-device combinations) \\
  & Execution order & Random (ensures unbiased KB population order; critical because RAG accuracy depends on prior profiling history) \\
  & KB lifecycle & Empty at start $\rightarrow$ grows with each Active Profiling execution $\rightarrow$ reset after batch completion (guarantees reproducibility across runs) \\
\midrule
\multirow{1}{*}{Ground Truth}
  & Monitored Hardware & 24 combinations for NVIDIA Jetson, 24 combinations for GPU desktop PC, 5 combinations for RPi5 \\
\bottomrule
\end{tabular}
\end{table*}

\subsection{Workload Composition}
\label{subsec:workload-composition}
53 evaluation workloads are tested and benchmarked.
Six datasets are used: CIFAR-10, FMNIST, MNIST, Forest Covertype, HIGGS, and Wine Quality (all pre-processed as per the dataset preparation pipeline). The sample and epoch counts are fixed for each device in order to reflect realistic edge constraints. These differ from device to device so that each device operates at full capacity while still being able to complete its task. 
JetsonThor is tracking unified memory, not RAM and VRAM. \gls{rpi} has no GPU (VRAM always 0.0\%). PC with GPU uses dedicated VRAM (with a total VRAM of 4,080~MB). Due to the constrained nature of the \gls{rpi} we could not benchmark all combinations on that hardware.

\subsection{Evaluation Metrics}
\label{subsec:evaluation-metrics}
For each prediction, \gls{mape} is computed as:
\begin{equation}
\text{MAPE} = \frac{|\hat{y} - y|}{|y|} \times 100\%
\label{eq:mape}
\end{equation}
where $\hat{y}$ is the predicted value and $y$ is the ground truth value. \gls{mape} is calculated per metric (Forecasting Time, CPU, GPU, RAM, VRAM, uRAM) and then averaged across all workloads for a given algorithm-dataset-device combination in the heatmaps which are here displayed for the \gls{rpi}, the PC with GPU and the Nvidia Jetson Thor.
Additionally all single combinations are plotted per data point in the scatter plots, where we combine the results for RAM and uRAM in one plot for increased comparability.

\subsection{Results}
\label{subsec:results}
\begin{figure*}
    \centering
    \includegraphics[width=1\linewidth]{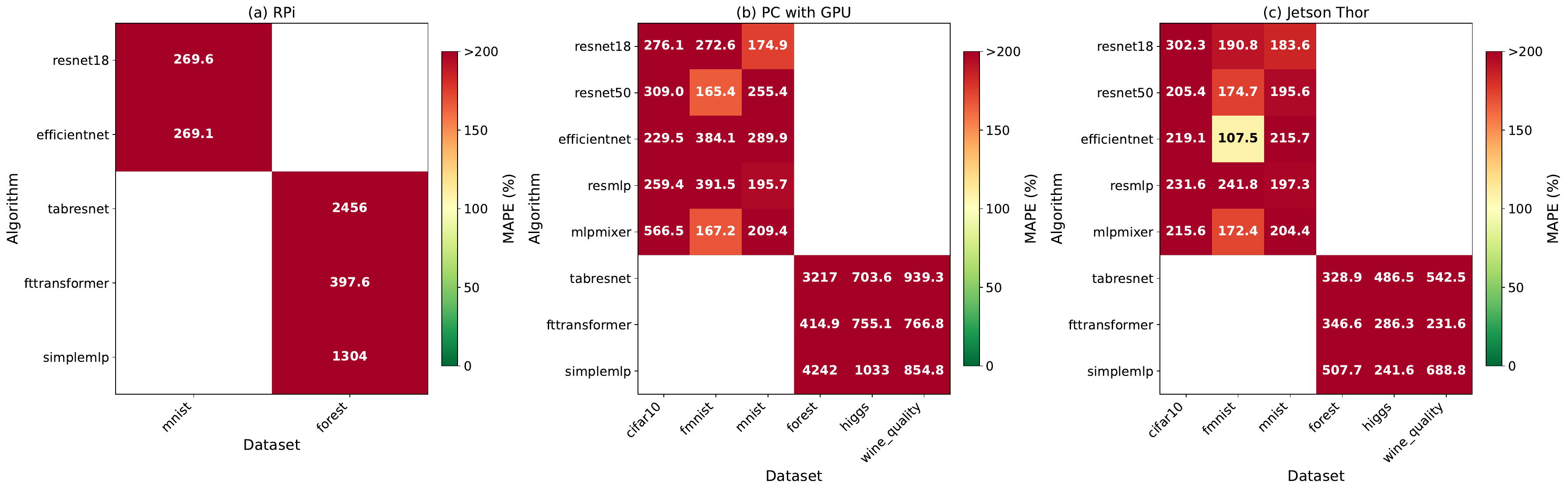}
    \caption{Overall prediction accuracy heatmap for the Zero-Shot LLM forecast}
    \label{fig:heatmap_zeroshot}
\end{figure*}

\begin{figure*}
    \centering
    \includegraphics[width=1\linewidth]{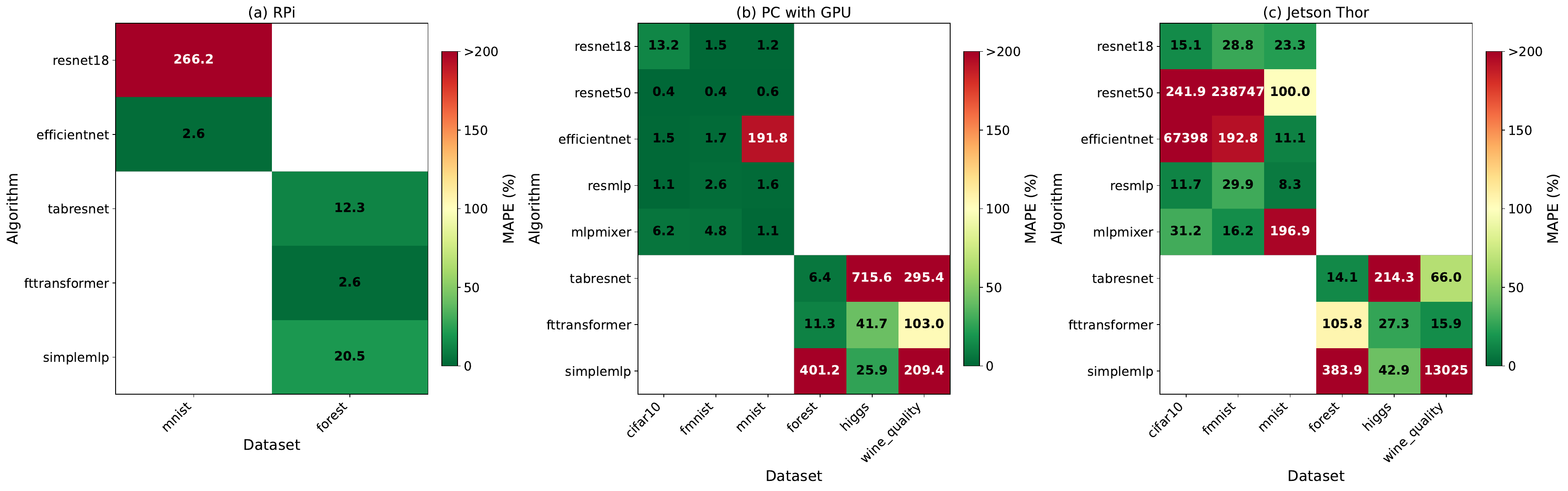}
    \caption{Overall prediction accuracy heatmap for the agentic framework}
    \label{fig:heatmeap_withtime}
\end{figure*}

\begin{figure*}
    \centering
    \includegraphics[width=1\linewidth]{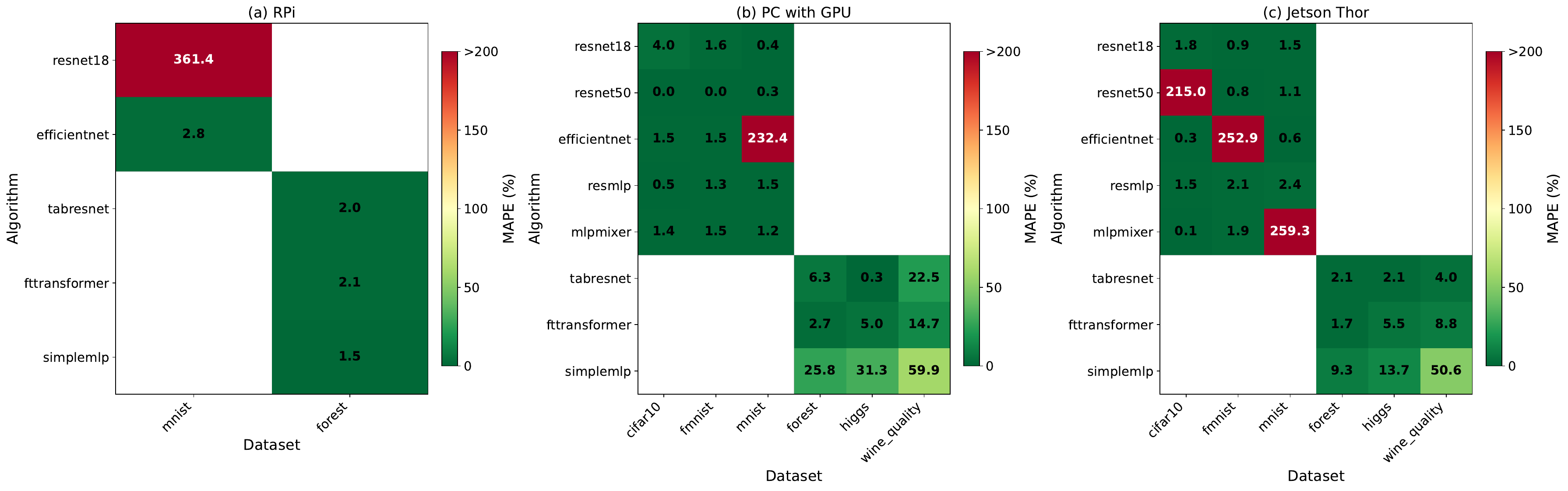}
    \caption{Overall prediction accuracy heatmap for the agentic framework without the training time prediction}
    \label{fig:heatmeap_notime}
\end{figure*}

\figurename~\ref{fig:heatmap_zeroshot}, \ref{fig:heatmeap_withtime}, and \ref{fig:heatmeap_notime} present the overall \gls{mape} across all predicted metrics for workloads executed on the \gls{rpi}, the PC with a GPU, and the NVIDIA Jetson Thor. The heatmaps compare the Zero-Shot LLM forecast against the proposed agentic framework, both with and without the inclusion of training time predictions.

As shown in \figurename~\ref{fig:heatmap_zeroshot}, the Zero-Shot LLM approach yields highly inaccurate predictions, with \gls{mape} values broadly exceeding 200\% and extreme outliers surpassing 4000\%. Across all devices, predictions for vision algorithms and datasets exhibit comparatively lower errors than tabular datasets and algorithms, though the absolute error remains prohibitively high. \figurename~\ref{fig:heatmeap_withtime} demonstrates that the agentic system significantly improves forecasting accuracy. In optimal cases, the \gls{mape} is reduced to approximately 0.4\%. However, a few outliers persist within this framework across various hardware-algorithm combinations.\figurename~\ref{fig:heatmeap_notime} displays the agentic framework's accuracy when training time predictions are excluded from the overall metric calculation. Under this condition, \gls{mape} drops to near 0\% for the vast majority of combinations. The few remaining outliers in this scenario correspond exclusively to edge cases where the system's knowledge base lacked any related workload data.

\begin{figure*}
    \centering
    \includegraphics[width=1\linewidth]{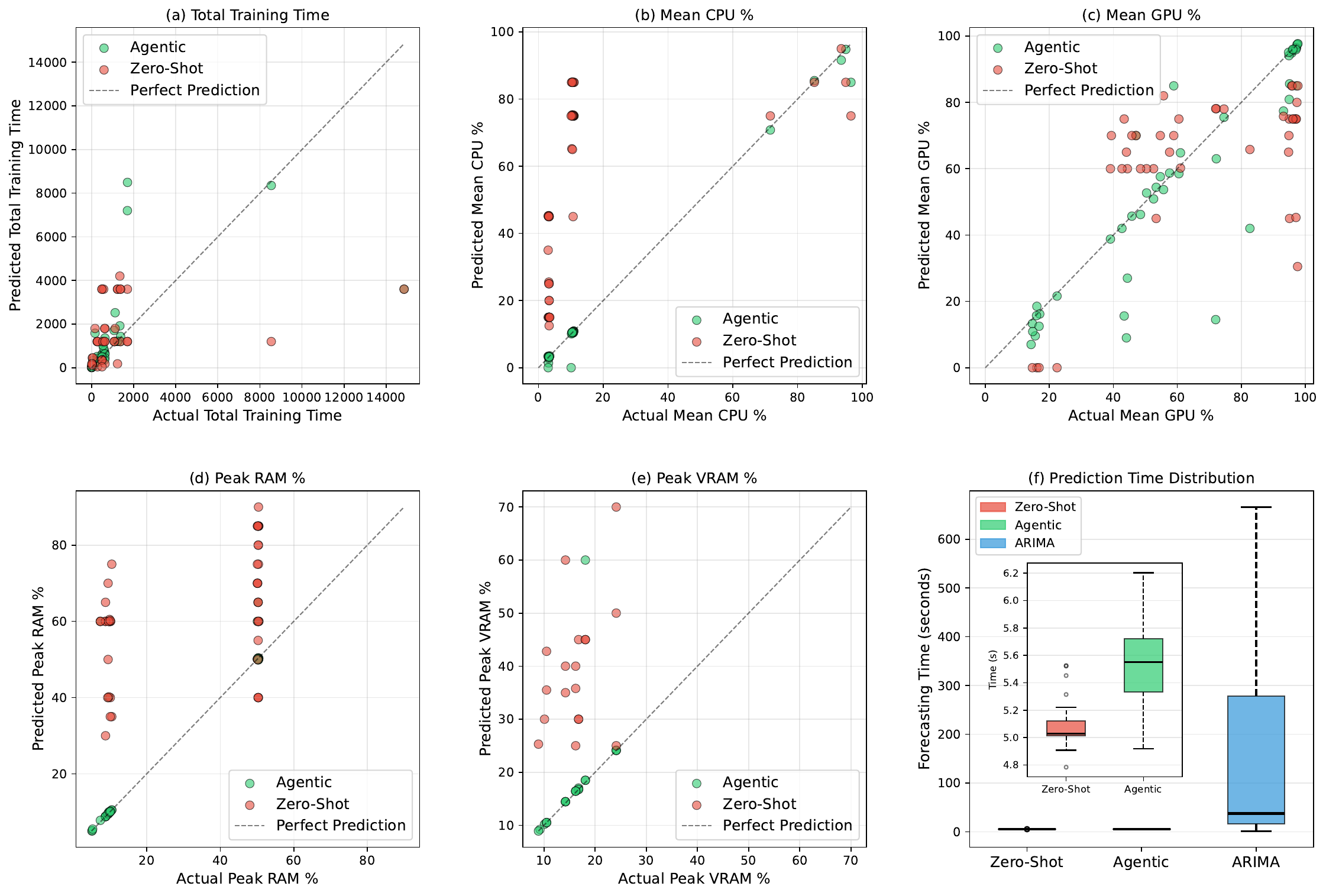}
    \caption{Predicted vs. Actual scatter plots for performance metrics (a-e) and Prediction Time Distribution (f).}
    \label{fig:scatterplot}
\end{figure*}

\figurename~\ref{fig:scatterplot} visualizes the predicted versus actual values for total training time, mean CPU usage, mean GPU usage, peak RAM, and peak VRAM, alongside the distribution of forecasting times. To maintain visibility in \figurename~\ref{fig:scatterplot}a, extreme outliers have been filtered out. For the remaining data points, the Agentic framework and Zero-Shot LLM show comparable spread around the actual training times, though the Agentic framework accounts for the broader \gls{mape} improvements noted previously. The divergence in accuracy becomes clearer in hardware utilization metrics. In \figurename~\ref{fig:scatterplot}b, the Agentic forecast closely tracks the perfect prediction line for mean CPU usage, whereas the Zero-Shot LLM consistently overestimates. A similar but less pronounced trend is visible for GPU usage (\figurename~\ref{fig:scatterplot}c), where the Agentic framework exhibits lower variance than the Zero-Shot method, albeit with a slight tendency to underestimate actual usage. For memory metrics, the Agentic framework achieves near-perfect alignment with actual peak RAM (\figurename~\ref{fig:scatterplot}d) and peak VRAM (\figurename~\ref{fig:scatterplot}e), while the Zero-Shot LLM generates scattered and significantly inflated predictions. \figurename~\ref{fig:scatterplot}f details the forecasting time required by each method. The Zero-Shot LLM is the fastest, averaging approximately 5.1 seconds. The Agentic framework requires slightly more time, averaging roughly 5.5 seconds. Both LLM-based approaches execute significantly faster and with vastly lower variance than the classical ARIMA baseline, which exhibits forecasting times ranging from tens of seconds up to over 600 seconds.

To evaluate the robustness of the Agentic framework, various hyperparameter and initial condition permutations were tested. Modifying the random seed for workload profiling yielded highly consistent outcomes, demonstrating algorithmic stability. However, the order of workload profiling measurably impacted both the knowledge base growth and prediction accuracy. Sorting workloads exclusively by device or by algorithm resulted in a smaller final knowledge base and the highest error rates for training time predictions. 

Additionally, we evaluated the impact of the k-NN distance threshold on the \gls{rag} routing logic. A low similarity threshold frequently bypassed the \gls{rag} system entirely, as workloads rarely met the strict similarity criteria. In contrast, an excessively high threshold allowed unrelated workloads to trigger the \gls{rag} pipeline, preventing the knowledge base from expanding with new ground truth data.

\begin{table*}
  \centering
  \caption{Hardware and System Specifications}
  \label{tab:hardware_specs}
  \begin{tabular}{@{}lllll@{}}
    \toprule
    \textbf{System / Device} & \textbf{CPU} & \textbf{GPU} & \textbf{RAM} & \textbf{Operating System} \\ \midrule
    Raspberry Pi 5 & Broadcom BCM2712 (Quad-core) & VideoCore VII & 16GB & Ubuntu Server 25.04 \\ 
    NVIDIA Jetson Thor & NVIDIA Grace (Arm Neoverse V2) & NVIDIA Blackwell & 128GB\textsuperscript{a} & Jetson Linux \\
    Custom Workstation & Intel i9-14900 (32) @ 5.500GHz & AMD Radeon RX 6400 & 32GB & NIXOS 24 \\
    \bottomrule
    \multicolumn{5}{l}{ \textsuperscript{a}{Unified Memory} }
  \end{tabular}
\end{table*}

\begin{table*}[h!]
\centering
\caption{Dataset Specifications}
\label{tab:dataset_specs}
\resizebox{\textwidth}{!}{%
\begin{tabular}{@{}lllrrrrllc@{}}
\toprule
\textbf{Algorithm} & \textbf{Dataset Name} & \textbf{Dataset type} & \textbf{Num Rows} & \textbf{Num Columns} & \textbf{Num Classes} & \textbf{Num Images} & \textbf{Image Size (Pixels)} & \textbf{Document Type}  & \textbf{Source} \\ 
\midrule

CNN & fashionMNIST & Images & - & - & 10 & 70000 & 28x28(greyscale) & idx3-ubyte & \cite{fmnistDataset}\\ 
    & MNIST &        & - & - & 10 & 70000 & 28x28 (greyscale) & idx3-ubyte & \cite{mnistDataset}\\ 
    & CIFAR-10 &        & - & - & 10 & 60000 & 32x32 (RGB) & binary & \cite{cifar10Dataset}\\ 

\addlinespace
MLP & wine\_quality & Numerical & 4900 & 12 & 10  & - & & csv & \cite{WineQuality}\\
    & HIGGS &           & $>$ 1 Mio & 29 & 2  & - & & csv & \cite{HiggsDataset} \\ 
    & forest\_dataset &           & 10.000 & 55 & 7 & - & & csv & \cite{forestdataset}\\ 
\bottomrule
\end{tabular}%
}
\end{table*}

\begin{table}[bhtp]
  \centering
  \caption{Overview of Selected AI Architectures}
  \label{tab:algorithm_specs_short}
  \begin{tabular}{@{}llc@{}}
  \toprule
  \textbf{Algorithm} & \textbf{Architecture Type} & \textbf{Approx. Parameters} \\ \midrule
  ResNet-18          & CNN & $\sim$11 Million \\ 
  ResNet-50          & CNN & $\sim$25.6 Million \\ 
  EfficientNet (B0)  & CNN & $\sim$5.3 Million \\ 
  Simple MLP         & Multi-Layer Perceptron & Variable \\ 
  ResMLP (S12)       & pure MLP & $\sim$15 Million \\ 
  MLP-Mixer (B/16)   & pure MLP & $\sim$59 Million \\ 
  Tabular Resnet     & Residual MLP (Tabular) & Variable ($\sim$0.1M – 2M) \\ 
  FTTransformer      & Attention-based (Tabular) & Variable ($\sim$1M – 5M)  \\ 
\bottomrule
\end{tabular}
\end{table}
\subsection{Discussion}
\label{subsec:discussion}

Our study reveals some limitations when using unaugmented LLMs for hardware-specific performance forecasting. The inaccuracy of the Zero-Shot LLM underscores that while LLMs possess broad conceptual knowledge of neural network architectures, they lack the intrinsic quantitative reasoning required to accurately estimate hardware-bound metrics without contextual grounding. It should be noted that our framework successfully addresses this issue by enriching the LLM with non-parametric memory via the \gls{rag} technique. By grounding the LLM's predictions in empirical data, the system successfully infers static or highly correlated metrics (such as memory footprint or parameter scaling) with near-perfect accuracy. 

The results also reveal that training time prediction is still a critical bottleneck for the agentic system. Execution time is a highly dynamic metric, sensitive to non-linear hardware-specific factors that an LLM struggles to extrapolate. Variables such as thermal throttling on edge devices (evident on the Jetson Thor and \gls{rpi}), memory bandwidth saturation, and I/O overhead do not scale linearly with dataset size or model parameters. Consequently, even when provided with closely related workloads via \gls{rag}, the LLM occasionally miscalculates temporal scaling factors, resulting in massive outliers.

Another limitation we observed in the experiments is that the system was susceptible to "cold start" scenarios. When \gls{rag} cannot retrieve knowledge about a specific domain or dataset, the agentic framework inevitably regresses toward Zero-Shot performance levels, explaining the minor isolated outliers that persist. This specific limitation can be addressed by adopting a hybrid architecture: offloading execution time and latency estimations to a dedicated deterministic regressor or an analytical hardware model, while preserving the agentic forecaster for resource prediction.

We also observe that the Agentic framework outperforms Zero-Shot LLM across all resource utilization metrics. This highlights the core advantage of the proposed architecture: the ability to gather new information and self-calibrate based on empirical ground truth. By passing similar workloads into the LLM's context window via \gls{rag}, the system anchors the LLM's inherently abstract hardware knowledge to factual, localized data. Most significantly, the accuracy does not come at the cost of operational delays. While the Agentic framework requires marginally more time than the Zero-Shot method due to the overhead of sequential LLM calls and database queries, it remains orders of magnitude faster than full monitoring or classical ARIMA forecasting. 

Despite the operational advantages of our agentic framework, we also acknowledge several architectural and experimental limitations. First, the full integration of a secure sandbox environment is exceptionally complex. As a result, the sandbox mechanism in this study is emulated by using the created and described dataset. Deploying a fully functional, isolated execution environment on constrained edge devices introduces significant system-level integration challenges that fall outside the scope of this study. Second, the reliance on active monitoring and telemetry inherently introduces computational noise into the collected ground-truth data. Although eliminating this interference entirely is not possible in real-world edge deployments, we mitigated its impact by carefully calibrating the frequency of monitoring calls to get an optimal balance between data precision and profiling overhead. 

In addition, the proposed system is currently agnostic to workload hyperparameters. Throughout our evaluation, we utilized a fixed set of hyperparameters to enable direct, equitable comparisons of AI workloads across heterogeneous hardware platforms. Exploring the vast space of possible hyperparameter combinations is computationally prohibitive and time-consuming, and doing so would ultimately prevent consistent, baseline comparisons of the underlying hardware performance.
Furthermore, the dataset in this study has not been subjected to comprehensive statistical evaluation, which is a challenge. 
Lastly, our modeling of CPU and GPU is rather simplified. Modern heterogeneous processors utilize highly complex microarchitecture features, such as dynamic voltage and frequency scaling and proprietary scheduling algorithms, which needs to be addressed in future work.

\section{Conclusion}
\label{sec:concl}

We proposed and evaluated a novel autonomous agentic framework designed to profile and allocate resources for zero-knowledge \gls{ai} workloads in highly heterogeneous and resource-constrained edge computing environments. By integrating a continuous self-calibration mechanism, our approach successfully addresses the fundamental challenges of operational drift and the absence of reliable ground truth typically encountered in open-ended, decentralized deployments.

The main contribution of our study is the self-calibrating AI Workload Agent capable of orchestrating four complementary profiling modules. We demonstrated that while zero-shot \gls{llm} estimation suffers from significant prediction errors (exceeding 200\% \gls{mape}) when evaluating unknown executables, combining hardware resource monitoring with ARIMA-based forecasting establishes a highly reliable ground truth. Crucially, our novel adaptive parameter search utilizing three leaping strategies allows the system to dynamically approximate full-scale resource footprints and terminate early. Our comprehensive evaluation on a new open benchmark dataset across diverse hardware platforms (Raspberry Pi 5, NVIDIA Jetson Thor, and a GPU workstation) validates the framework's efficacy. By synthesizing empirical telemetry with a \gls{rag}-enhanced estimation module, the system successfully refines its own knowledge base, reducing prediction errors to single-digit \gls{mape} for well-covered workload classes. Furthermore, the proposed ARIMA leaping algorithm proved 52\% faster than classical models while maintaining equivalent accuracy.

Ultimately, this self-calibrating architecture provides a critical foundation for deploying reliable agentic edge intelligence. To further mature this framework, future research must address several key operational and architectural challenges. First, we will investigate comparing the ARIMA-based module with lightweight, on-device sequential neural networks (such as LSTMs or state-space models) to better capture non-linear resource degradation and transient microbursts. Because forecasting training time remains a significant bottleneck for agentic architectures, we also plan to integrate dedicated neural network-based runtime predictors. Furthermore, the system can be improved by optimizing the initial seeding strategy and implement dynamic, self adjusting k-NN thresholds to ensure robust \gls{rag} retrieval across varying edge computing environments. To complement this and to  ensure long-term scalability on constrained edge orchestrators, future iterations must implement dynamic memory management and eviction policies to prevent unbounded \gls{rag} knowledge base growth. We propose to extend the single-agent orchestrator into a fully decentralized, multi-agent cooperative system.

\bibliographystyle{IEEEtran}
\bibliography{references}
\end{document}